\documentclass[sigconf, natbib=true]{acmart}
\usepackage[utf8]{inputenc}
\newcommand{\header}[1]{\vspace*{1mm}\noindent\textbf{#1.}}

\usepackage{graphicx}
\usepackage{subcaption}  
\usepackage{threeparttable}
\usepackage{enumerate}
\usepackage[inline]{enumitem}
\usepackage{float}
\usepackage{lipsum} 
\usepackage[utf8]{inputenc}
\usepackage{float}
\usepackage{amsmath} 
\usepackage{booktabs} 
\usepackage[english]{babel}
\usepackage[T1]{fontenc}
\usepackage{multirow}
\usepackage{xspace}
\usepackage{siunitx}
\usepackage{hyperref}  
\AtBeginDocument{%
  \providecommand\BibTeX{{%
    \normalfont B\kern-0.5em{\scshape i\kern-0.25em b}\kern-0.8em\TeX}}}

\usepackage{threeparttable}
\usepackage{float}
\usepackage{setspace}

\AtBeginDocument{%
  \providecommand\BibTeX{{%
    Bib\TeX}}}

\copyrightyear{2025}
\acmYear{2025}
\setcopyright{cc}
\setcctype{by}
\acmConference[SIGIR '25]{Proceedings of the 48th International ACM SIGIR Conference on Research and Development in Information Retrieval}{July 13--18, 2025}{Padua, Italy}
\acmBooktitle{Proceedings of the 48th International ACM SIGIR Conference on Research and Development in Information Retrieval (SIGIR '25), July 13--18, 2025, Padua, Italy}\acmDOI{10.1145/3726302.3730285}
\acmISBN{979-8-4007-1592-1/2025/07}

\begin{document}

\title{Reproducibility, Replicability, and Insights into Visual Document Retrieval with Late Interaction}

\author{Jingfen Qiao}\authornote{Corresponding author.}
\affiliation{%
  \institution{University of Amsterdam}
  \city{Amsterdam}
  \country{The Netherlands}
}

\author{Jia-Huei Ju}
\affiliation{%
  \institution{University of Amsterdam}
  \city{Amsterdam}
  \country{The Netherlands}
  }

\author{Xinyu Ma}
\affiliation{%
  \institution{Baidu Inc.}
  \city{Beijing}
  \country{China}
}

\author{Evangelos Kanoulas}
\affiliation{%
  \institution{University of Amsterdam}
  \city{Amsterdam}
  \country{The Netherlands}
}

\author{Andrew Yates}
\affiliation{%
 \institution{Johns Hopkins University, HLTCOE}
 \city{Baltimore}
 \country{United States}
 }

\renewcommand{\shortauthors}{Jingfen Qiao, Jia-Huei Ju, Xinyu Ma, Evangelos Kanoulas, and Andrew Yates}

\begin{abstract}

Visual Document Retrieval (VDR) is an emerging research area that focuses on encoding and retrieving document images directly, bypassing the dependence on Optical Character Recognition (OCR) for document search. A recent advance in VDR was introduced by ColPali, which significantly improved retrieval effectiveness through a late interaction mechanism. ColPali's approach demonstrated substantial performance gains over existing baselines that do not use late interaction on an established benchmark. In this study, we investigate the reproducibility and replicability of VDR methods with and without late interaction mechanisms by systematically evaluating their performance across multiple pre-trained vision-language models. Our findings confirm that late interaction yields considerable improvements in retrieval effectiveness; however, it also introduces computational inefficiencies during inference. Additionally, we examine the adaptability of VDR models to textual inputs and assess their robustness across text-intensive datasets within the proposed benchmark, particularly when scaling the indexing mechanism. Furthermore, our research investigates the specific contributions of late interaction by looking into query-patch matching in the context of visual document retrieval. We find that although query tokens cannot explicitly match image patches as in the text retrieval scenario, they tend to match the patch contains visually similar tokens or their surrounding patches.

\end{abstract}



\maketitle
\section{Introduction}
Vision-language model (VLM) pre-training~\cite{radford2021learning, li2022blip, chen2023pali, zhai2023sigmoid} has facilitated the unification of textual and visual modalities, enhancing the applicability of document retrieval in real-world scenarios~\cite{ma2024visa, yang2023atomic}. 
For instance, M-BEIR~\cite{wei2024unir} has established a benchmark including various document retrieval tasks involving text, images, or both.
However, these settings are not always practical, as real-world documents are often unstructured and not optimally preprocessed~\cite{li2024readocunifiedbenchmarkrealistic}. Instead, they may be more readily available in image-based formats, such as screenshots.
%
Extracting meaningful content from such visual documents typically requires Optical Character Recognition (OCR) to convert visual text into machine-readable format~\cite{lin2022retrieval}. Additionally, the parsing pipeline often employs multiple detection modules to extract figures, tables, and other embedded content. 
This introduces complexity and leads to information loss~\cite{cho2024m3docragmultimodalretrievalneed}, particularly for web pages with spatial layouts~\cite{li2023enhancing}, or positional relationships with visual elements~\cite{wang2024docllm}.

To address these challenges, recent research has introduced a novel paradigm in document retrieval---Visual Document Retrieval (VDR)---which treats documents as multimodal images rather than purely textual entities.
VDR offers a more natural fit for real-world document storage and retrieval, particularly in domains where documents are predominantly compiled into PDF files, such as financial reports~\cite{Zhu_2022}, government documents~\cite{Zhu_2022} and presentation slides~\cite{tanaka2023slide}. 
By retrieving documents directly in their native image format, VDR bypasses the preprocessing pipeline while preserving the inherent multimodal structure of the embedded content.
However, VDR introduces new challenges of visual understanding in the encoding process, such as diverse and unstructured contexts spread throughout an image, emphasizing the critical need for more fine-grained contextualized multi-modal representations.

Recent studies by~\citet{ma2024unifyingmultimodalretrievaldocument} and~\citet{faysse2024colpali} have shown that large vision-language models (LVLMs) can effectively represent visual documents by encoding them as sequences of image patches~\cite{dosovitskiy2021an}. 
Notably, LVLMs have achieved success similar to that of decoder-only language models~\cite{brown2020languagemodelsfewshotlearners} by treating patches as context for next token prediction, thereby enabling more fine-grained contextualization among the image.
Such contextualized representations are informative and can benefit in the context of visual document retrieval. 
For example, DSE~\cite{ma2024unifyingmultimodalretrievaldocument} uses a CLIP-based~\cite{radford2021learning} vision encoder to encode images into patch embeddings, which are then fed into a LVLM (e.g., Phi-3~\cite{abdin2024phi3technical}) to generate the contextualized patch embeddings and further compressed into a single representation of visual document.
This representation can be seamlessly integrated into dense retrieval approaches with simple vector operation to estimate more accurate relevance scores.

Unlike the single-vector approach used in DSE, ColPali~\cite{faysse2024colpali}, a novel approach for visual document retrieval, integrates the \emph{late-interaction} mechanism~\cite{khattab2020colbert} into relevance score estimation.
Specifically, the mechanism computes the similarity between each query token and each image patch embeddings, and then aggregate all similarity scores into the final relevance score of the given visual document.
Empirical study of ColPali demonstrates that this mechanism significantly outperforms various baseline methods, highlighting its advantage in VDR. Late interaction, also known as the multi-vector approach, is generally more expressive than the single-vector method because it captures both lexical and semantic matching in text retrieval~\cite{khattab2020colbert, wang2023repro}. However, how this matching extends to VDR, specifically the interaction between query tokens and image patches, is underexplored.

\begin{figure}
    \centering
    \includegraphics[width=0.85\linewidth]{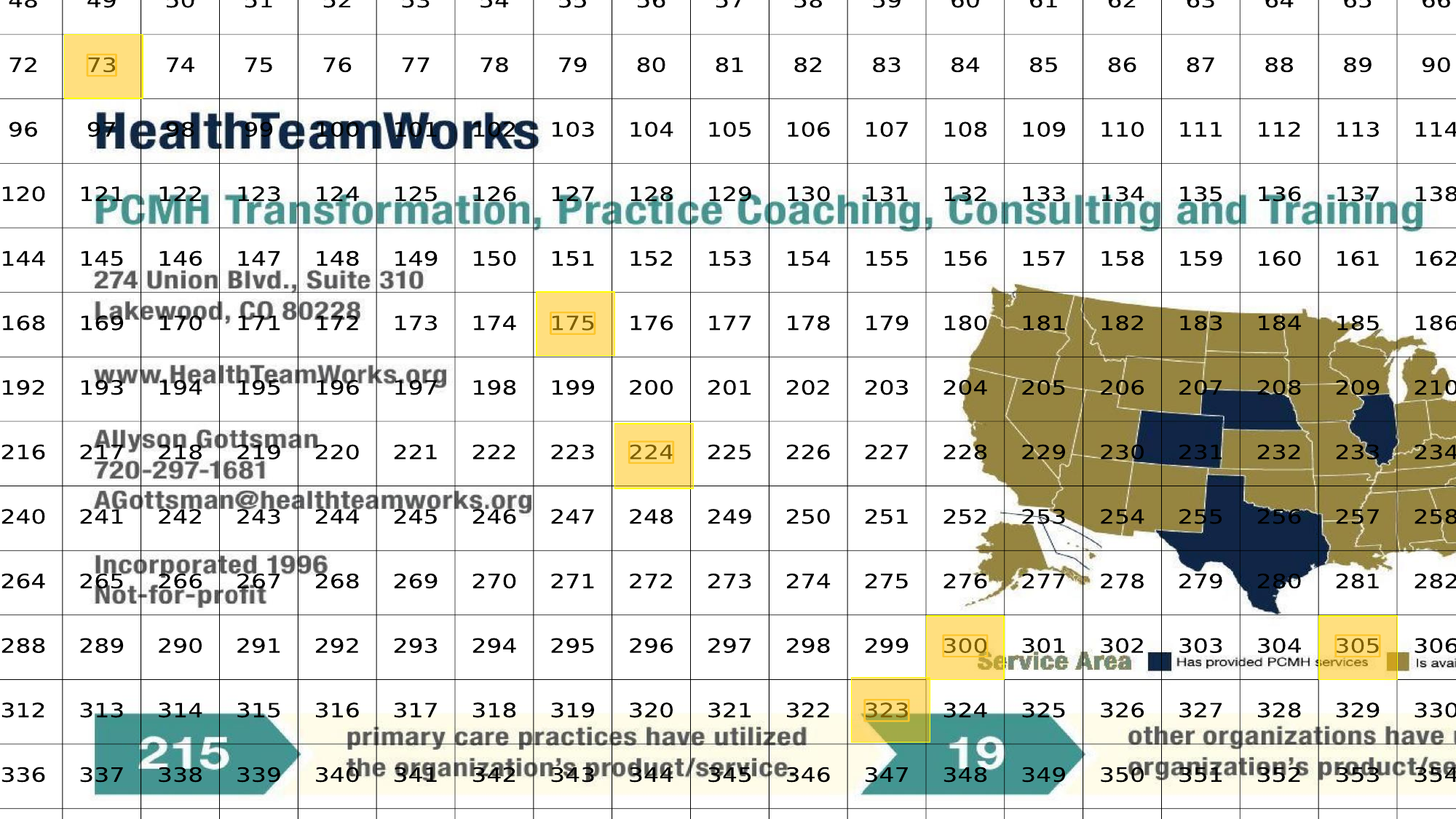}
    \caption{Example of late interaction matching between query and patch tokens; The matching document patch is highlighted by yellow. \textit{Query: What Services does Health Team Works Provide?}}
    \label{fig: 1}
\end{figure}

To address this gap, first, we reproduce ColPali training and demonstrate the strength of late interaction.
Second, we investigate ColPali's robustness under different document retrieval scenarios, including the zero-shot settings and the larger corpus settings.
Finally, we investigate ColPali's query-matching behavior.
Our findings can be categorized into the following parts: \textbf{Reproducibility:} We follow the original implementation and found: (i) we can completely reproduce ColPali training and achieve results similar to the reported ones; (ii) the improvement gap between ColPali and baseline settings are also reproducible and consistent with the claim in the original paper. \textbf{Replicability:} We identify the additional advantages of visual document retrieval in two practical scenarios:
(iii) training on visual documents can generalize to text document retrieval under zero-shot settings; (iv) ColPali is more robust to increased corpus size compared to OCR-based document retrieval, especially if the document contains higher non-textual portion (coverage) in the document, such as ArXiVQA. \textbf{Insights:} Our detail analysis discloses critical findings on query matching behavior: (v) visual document retrieval effectiveness is significantly correlated to text coverage in the image, and (vi) ColPali encourages more abstract lexical matching, as illustrated in Figure~\ref{fig: 1}, unlike text retrieval, which is dominated by lexical matching (exact same tokens in the document).  
We believe these findings will help explain the matching behavior of token-patch late interaction, identifying possible directions toward more fine-grained visual document retrieval design. The code is also available on GitHub.\footnote{https://github.com/JingfenQiao/ColPali-Reproducibility}

\section{Preliminaries}
In this section, we first outline the visual document retrieval (VDR) task. 
Next, we describe the background and detail the method we reproduced, focusing on the core techniques proposed in ColPali~\cite{faysse2024colpali}.

\subsection{Visual Document Retrieval}
\header{Task Definition and Notation}
The major difference between visual document retrieval (VDR) and ad-hoc document retrieval lies in the modality of the documents. 
Given a query text $Q=[q_1, q_2, ..., q_{|Q|}]$, the goal of VDR system is to retrieve a ranked list of documents from the collection $D\in\mathcal{C}$, where each document $D$ contains mixed modalities, such as texts, charts, tables, or diagrams.
Note that each document is stored in the image format, containing standalone information about a page or a particular screenshot.
For clarity, ``document'' and ``visual document'' are interchangeably and refer to the retrieval unit, but it would be processed as an image and further encoded into a sequence of patches $D=[d_1, d_2, ..., d_{|D|}]$.

\header{Benchmark Datasets}
Our experiments are on ViDoRe\footnote{\url{https://huggingface.co/vidore}}~\cite{faysse2024colpali} a comprehensive VDR benchmark proposed along with ColPali.
The benchmark comprises 10 domain-specific datasets. 
For example, as illustrated in Figure~\ref{fig: 3}, the visual document contains multiple embedded contents in Healthcare domain, such as title, texts and figures.
The testing queries are generated by Claude-3 Sonnet and further filtered based on human judgments.

\begin{table*}
    \centering
    \setlength\tabcolsep{6pt}
    \caption{Evaluation of ViDoRe benchmark datasets in terms of nDCG@5. 
    LI refers to models with late interaction. Values in \textcolor{red}{red} represent the improvement in performance when compare to single-vector variant.}
    \resizebox{.98\textwidth}{!}{
    \begin{tabular}{lllllllllllllr}
    \toprule
    \toprule
 Models& LI & arXiVQ& DocQ& InfoQ& TabF& TaTq& Shift& AI& Energy & Gov& Health&Avg. \\ 
 \midrule
\multicolumn{13}{c}{\textit{Reported Results}}\\ 
        \midrule
         BiPali &  no&  56.5&  30.0&  67.4&  76.9&  33.4&  43.7&  71.2&  61.9& 73.8& 73.6&58.8 \\
         ColPali &  yes & 83.0\textcolor{red}{\tiny{+26.5}} &58.5\textcolor{red}{\tiny{+28.5}}& 85.7\textcolor{red}{\tiny{+18.3}}& 87.4\textcolor{red}{\tiny{+10.5}}& 70.4\textcolor{red}{\tiny{+37}}& 77.4\textcolor{red}{\tiny{+33.7}}&  97.4\textcolor{red}{\tiny{+26.2}}&95.4\textcolor{red}{\tiny{+33.5.7}}& 96.2\textcolor{red}{\tiny{+22.4}}& 96.9\textcolor{red}{\tiny{+23.3}}&84.8\textcolor{red}{\tiny{+26}} \\
         ColQwen2 & yes & 88.0& 60.6& 92.4& 89.5& 81.5& 81.5& 99.4& 96.3& 96.3& 98.5&88.4 \\
\midrule
\multicolumn{13}{c}{\textit{Our Reproduced Results}}\\ 
        \midrule
         BiPali&  no&  54.4&  35.3&  61.8&  73.5&  31.8&  32.2&  65.1&  66.1& 66.4& 72.9&55.9 \\
         ColPali&  yes&  82.2\textcolor{red}{\tiny{+27.7}}&  58.0\textcolor{red}{\tiny{+22.7}}& 83.2\textcolor{red}{\tiny{+21.4}}&  84.1\textcolor{red}{\tiny{+10.6}}& 67.8\textcolor{red}{\tiny{+36}}&  76.0\textcolor{red}{\tiny{+43.8}}& 97.0\textcolor{red}{\tiny{+21.4}}&  94.7\textcolor{red}{\tiny{+28.6}}& 92.6\textcolor{red}{\tiny{+26.2}}& 94.0\textcolor{red}{\tiny{+21.2}}& 83.0\textcolor{red}{\tiny{+27.1}}\\
         BiQwen2&  no&  65.0 & 37.8&  68.1&  76.3&  38.3&  46.8&  72.3&  67.2& 74.6& 78.5&  62.5 \\
         ColQwen2&  yes&  83.6\textcolor{red}{\tiny{+18.6}} & 62.0 \textcolor{red}{\tiny{+24.2}}& 90.9\textcolor{red}{\tiny{+22.8}}&  88.7\textcolor{red}{\tiny{+12.4}}& 78.1\textcolor{red}{\tiny{+20.2}}&  86.3\textcolor{red}{\tiny{+39.8}}& 98.9\textcolor{red}{\tiny{+26.6}}&  95.4\textcolor{red}{\tiny{+28.2}}& 95.5\textcolor{red}{\tiny{+20.9}}& 97.5\textcolor{red}{\tiny{+24.6}}& 87.7\textcolor{red}{\tiny{+25.2}}\\ 

\bottomrule
\bottomrule
    \end{tabular}
    }
    \label{tab: table 1}
\end{table*}

\subsection{Reproduced Method: ColPali}\label{sec:preliminary}

\header{Bi-Encoder Architectures}
Bi-encoder architectures are widely adopted in dense retrieval approaches~\cite{li2022blip, radford2021learning, zhai2023sigmoid}, as they decouple the query and document encoding processes, allowing documents to be encoded offline and stored.
Typically, the relevance score is calculated as:
\begin{equation}
    s_{Q, D} = sim \big( {\rm pool}(E_{Q}), {\rm pool}(E_{D}) \big), 
    \label{eq:single}
\end{equation}
where $E_Q \in \mathbb{R}^{|Q|\times h}$ and $E_D\in \mathbb{R}^{|D|\times h}$ are the $h$-dimensional token embeddings of the query $Q$ and the document $D$, respectively.
The pooling operator, ${\rm pool}(\cdot)$, can be average pooling over token embeddings~\cite{reimers2019sentencebertsentenceembeddingsusing}, or an indicator of special token embedding (e.g., \texttt{[CLS]})~\cite{karpukhin2020densepassageretrievalopendomain}.
This operation compresses the query and document into vectors, enabling the relevance score to be estimated using function $sim()$, such as cosine similarity.
Note that in the context of VDR, the document encoder is a vision encoder from contrastive VLMs, such as CLIP~\cite{radford2021learning} or SigLIP~\cite{zhai2023sigmoid}. Thus, the length of the document tokens $|D|$ is the number of patches in the visual document.

\header{Contextualized Patch Embeddings}
Recently, many studies have observed that traditional contrastive pre-trained vision encoders struggle to capture contextualized information within the image~\cite{lewis2024clip, ma2024unifying}, resulting in suboptimal representation and thereby hindering downstream applications.
To enhance visual understanding, ColPali leverages PaliGemma~\cite{beyer2024paligemma}, a large vision-language model (LVLM) to further contextualize the patch embeddings derived from the vision encoder, SigLIP~\cite{zhai2023sigmoid}. 
Similarly,~\citet{ma2024unifyingmultimodalretrievaldocument} adopt the LVLM, Phi-3-Vision~\cite{abdin2024phi3technical} with the CLIP vision encoder to enrich the context of patch embeddings.
Many recent LVLMs have also explored advanced enhancement like dynamic tokenization~\cite{Qwen2VL}, which further improve visual understanding capability of VDR's encoder.

\header{Token-Patch Late Interaction}
To fully leverage contextualized patch embeddings, ColPali adopts the \emph{late-interaction} mechanism proposed in ColBERT~\cite{khattab2020colbert}, leading to significant improvements in VDR performance.
Unlike the single-vector dense retrieval like Eq.~\eqref{eq:single}, ColPali computes all token-patch interactions to capture fine-grained matching between the query $Q$ and document $D$ as:
\begin{align}
    s_{Q, D} = \sum_{i=1}^{|Q|}\max_{j=1,...,|D|} sim( E_{q_i}, E_{d_j} ), \label{eq:multi}
\end{align}
where $E_{q_i}, E_{d_j} \in \mathbb{R}^{1\times h}$ represents the contextualized embeddings of the $i$-th query-token and the $j$-th document-patch, generated from LVLMs.
The relevance score is estimated by summing the maximum similarity scores across patches for each query token.
Note that ColPali follows ColBERT query augmentation design by right-padding query input with special token ``\texttt{<|endoftext|>}'' to maximum length.
Regarding the fine-tuning, ColPali also follows the ColBERT-v1 and fine-tunes with pairwise contrastive learning using in-batch negative, formulated as:
\begin{equation}
    \mathcal{L}(Q, D^+) = 
    -\log \dfrac{ \exp (S_{Q, D^+}) /\tau }{ \exp (S_{Q, D^+} /\tau) + \exp  (S_{Q, D^-} /\tau )},
    \label{eq:loss}
\end{equation}
where $S_{Q, D^-}$ denotes the maximum relevance score among all the other in-batch negative documents. This loss encourages the model to distinguish the positive document $D^+$ and the most challenging negative $D^-$ in the mini-batch.

\begin{table*}
    \centering
        \caption{Evaluation results of retrieval effectiveness (nDCG@5) using OCR text and screenshot images as document input.}

    \setlength\tabcolsep{6pt}
    \resizebox{.98\textwidth}{!}{
    \begin{tabular}{lllccclllllccc} 
    \toprule
    \toprule
         Model &  Size& Doc Format &  arXiVQ&  DocQ&  InfoQ&   TabF& TaTq& Shift& AI& Energy &Gov&  Health&  Avg. \\ 
    \midrule
         \multicolumn{14}{c}{\textit{Zero-shot on visual document domain}}\\ 
         \midrule
         Jina\_ColbertV2& 559M  &Text (OCR)&  33.5&  41.5&  67.7&   88.1& 34.0& 57.6& 81.4& 70.4&77.5&  79.6&  63.1 \\ 
    
         Colbert-V2&   110M&Text (OCR)&  34.8&  39.5&  64.5&   58.1& 32.6& 37.7& 77.8& 74.2&76.2&  76.9&  57.2 \\ 
    
         BGE-M3&   100M&Text (OCR)&  33.4&  36.4&  67.2&   90.6& 41.5& 71.8& 91.9& 86.2&87.9&  87.4&  69.4 \\ 
    
         GTE-Qwen2&  1.78B&Text (OCR)& 40.3& 30.6& 70.8& 82.7& 43.0& 71.4& 87.3& 84.6& 88.2& 92.4&69.1 \\ 
 GTE-Qwen2& 7.72B& Text (OCR)& 40.7& 35.2& 76.2& 93.2& 44.6& 72.6& 89.6& 88.0& 88.0& 93.2&72.1\\ 
 Qwen2-VL& 2.21B& Text (OCR)& 22.9& 9.9& 35.4& 35.0& 10.4& 32.4& 33.1& 37.8& 39.1& 44.9&30.1\\
     \midrule
         \multicolumn{14}{c}{\textit{Fine-tuning on visual document domain}}\\ 
         \midrule
    
        DSE--Qwen2&  2.21B&Text (OCR)& 44.3& 43.3& 77.2& 88.3& 66.2& 74.4& 96.8& 90.3& 93.1& 95.4&76.9 \\
    
        DSE--Qwen2&  2.21B&Image    & 85.4& 56.0& 87.8& 93.8& 69.1& 80.5& 98.4& 91.9& 95.8& 96.8&85.5 \\
    
        BiQwen2&   2.21B&Text (OCR)&  23.4&  17.6&  35.7&   49.5& 23.7& 12.9& 55.8& 44.7&41.0&  43.2&  34.7 \\ 
    
        BiQwen2&   2.21B&Image    &  65.0&  37.8&  68.1&   76.3& 38.3& 46.8& 72.3& 67.2&74.6&  78.5&  62.5 \\ 
    
        ColQwen2&  2.21B&Text (OCR)& 38.2& 47.4& 72.0& 82.4& 75.9& 81.3& 98.0& 92.6& 92.0& 95.1&77.5 \\
    
        ColQwen2&  2.21B&Image    & 86.0& 60.9& 91.4& 89.3& 79.0& 85.7& 99.0& 95.3& 94.9& 98.0&88.0 \\
    \bottomrule
    \bottomrule
    \end{tabular}
    }
    \label{tab: table 2}
\end{table*}

\section{Reproducibility}
In this section, we follow the official implementation\footnote{\url{https://github.com/illuin-tech/colpali}} released by original ColPali team and reproduce both the training and evaluation.
In \textbf{RQ1.1}, we validate the effectiveness of late interaction design. In \textbf{RQ1.2}, we reproduce the identical training on single-vector settings, aiming to verify the necessity of late interaction.

\vspace{1em}\noindent
\textbf{RQ1.1: Can we completely reproduce ColPali and achieve the same effectiveness on visual document retrieval?}
To assess the effectiveness of late interaction in the visual document retrieval (VDR) tasks, we conduct experiments with the same multi-vector settings and fine-tune the models from scratch using Google PaliGemma~\cite{beyer2024paligemma}\footnote{\url{https://huggingface.co/google/paligemma-3b-mix-448}} as initialization with the compatible vision encoder, SigLIP~\cite{alabdulmohsin2024gettingvitshapescaling}\footnote{\url{https://huggingface.co/google/siglip-so400m-patch14-384}}.
Our reproduction efforts strictly follows the hyperparameters and training procedures specified in the original paper.
In addition to our reproductions, we also re-evaluate the author-provided checkpoints: ColPali\footnote{\url{https://huggingface.co/vidore/colpali-v1.3}} and the ColQwen2-v1.0.\footnote{\url{https://huggingface.co/vidore/colqwen2-v1.0}}

\paragraph{Experimental Setups} 
Consistent with the original paper~\cite{faysse2024colpali}, our reproduction uses the same training dataset,\footnote{\url{https://huggingface.co/datasets/vidore/colpali\_train\_set}} which has 127,460 query-document pairs. Of these, 63\% were sourced from publicly available academic datasets and 37\% were derived from a synthetic dataset. Each model was trained for 3 epochs, matching the duration used in ColPali checkpoint. 
We use the same in-batch contrastive objective as formulated in Eq.~\eqref{eq:loss}.
Training was performed using four NVIDIA A6000 GPUs with a batch size of 128 and gradient accumulation steps of 2, simulating the original training performed on 8 MI250X AMD GPUs with a total batch size of 256. Low-rank adapters (with parameters \(\alpha =32\) and \(r = 32\)) were integrated into the transformer layers of the large vision-language model, along with a randomly initialized final projection layer. The learning rate is set as \num{5e - 5} with a linear decay strategy and a warm-up phase covering \(2.5\%\) of the training duration to gradually acclimate the model. 
Finally, we assess whether this training process can be applied to different backbone and conduct the same training on another LVLM: Qwen2-VL~\cite{Qwen2VL, Qwen-VL}.\footnote{\url{https://huggingface.co/Qwen/Qwen2-VL-2B-Instruct}}

\paragraph{Results} Table~\ref{tab: table 1} presents the evaluation results of both the reported and the reproduced results on the ViDoRe benchmark. 
The performance of our reproduced ColPali and ColQwen2 models (the second block in Table~\ref{tab: table 1}) is slightly lower than the results obtained from re-evaluating the author-provided checkpoints (the first block). However, we consider this discrepancy is reasonable and can be primarily attributed to differences in GPU architectures and hardware configurations. 
Our reproduction also showcases the clear performance gain when using Qwen2-VL as the backbone model, where the reproduced ColQwen2 outperforms our reproduced ColPali by 4.7 points of nDCG@5. 
This observation aligns with the finding stated in the original paper: stronger backbone models lead to more effective visual document retrieval effectiveness.

\paragraph{Answer}
Our results demonstrate that the training process for ColPali is indeed reproducible. 
We also observe similar gains by upgrading LVLM backbone from PaliGemma to Qwen2-VL, as shown in the gaps between ColPali and ColQwen2 in the first two blocks of Table~\ref{tab: table 1}.
With the confirmed reproducible training process of ColPali, we further conduct experiments to validate the benefit of the design compared to the single-vector settings.

\begin{figure*}[h]
    \centering 
    \begin{subfigure}{0.33\textwidth}
        \includegraphics[width=\linewidth]{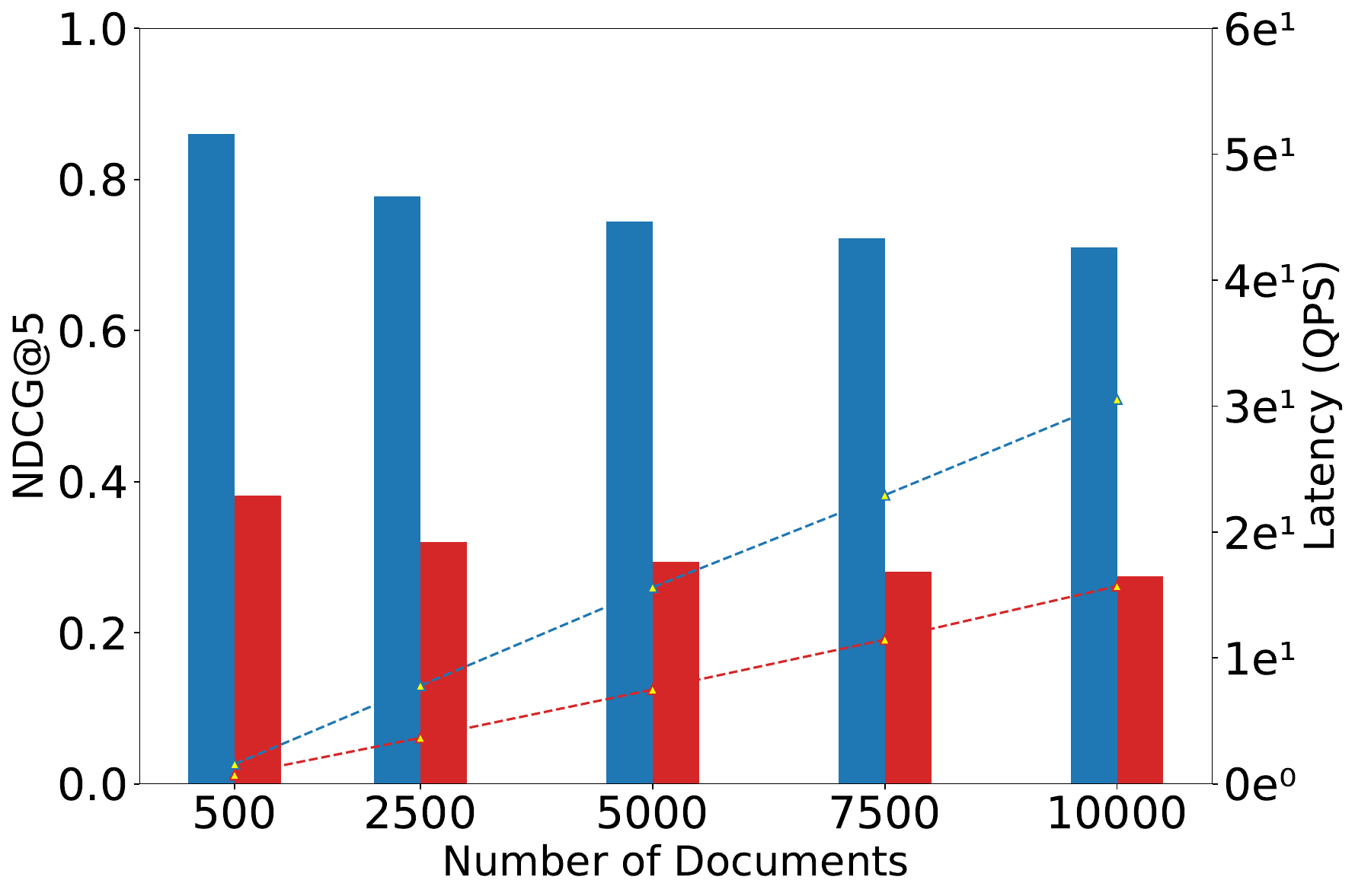}
        \caption{ColQwen2 Evaluation on arXiVQA}
    \end{subfigure}
    \hfill 
    \begin{subfigure}{0.33\textwidth}
        \includegraphics[width=\linewidth]{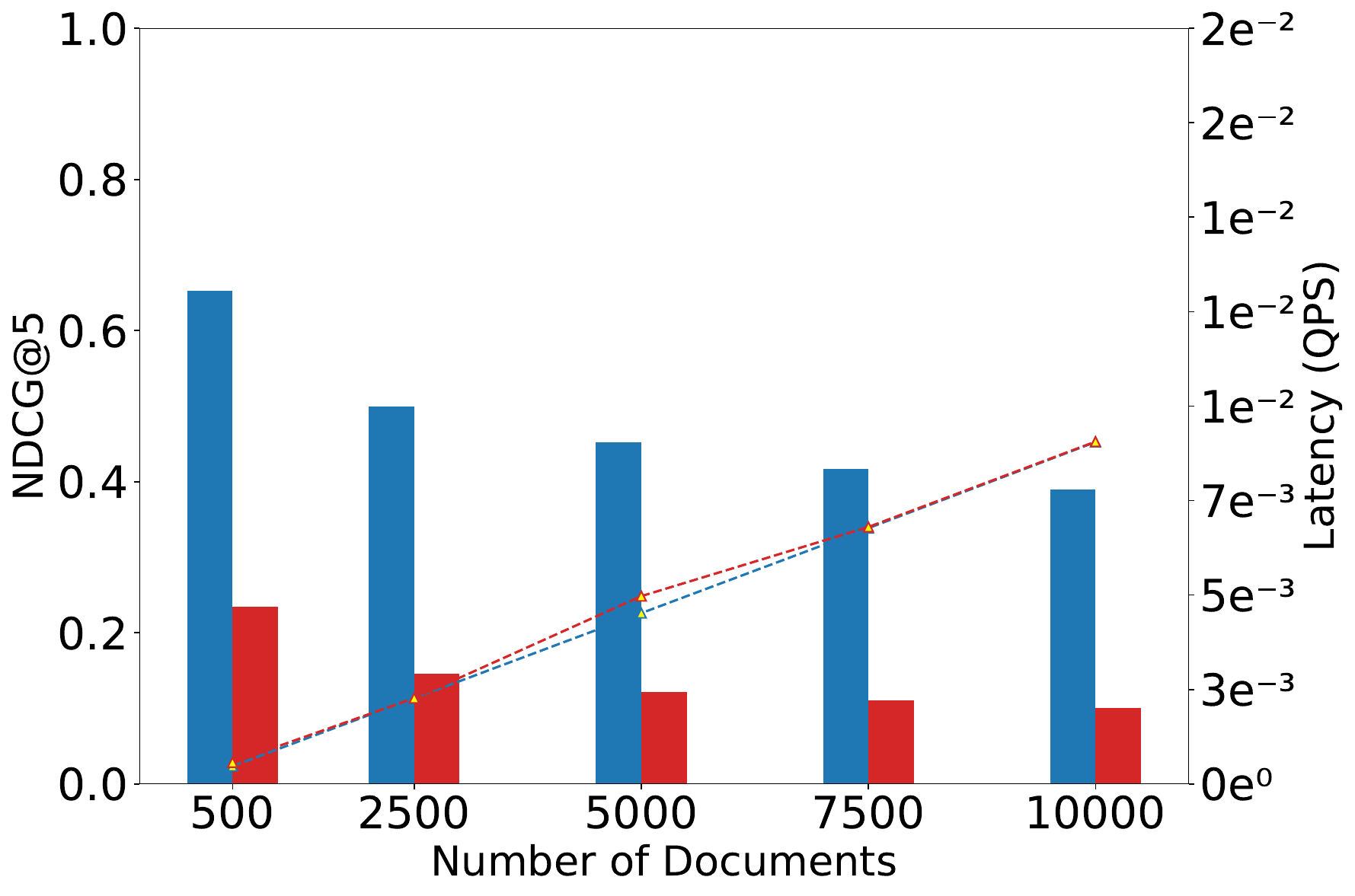}
        \caption{BiQwen2 Evaluation on arXiVQA}
    \end{subfigure}
    \hfill 
    \begin{subfigure}{0.33\textwidth}
        \includegraphics[width=\linewidth]{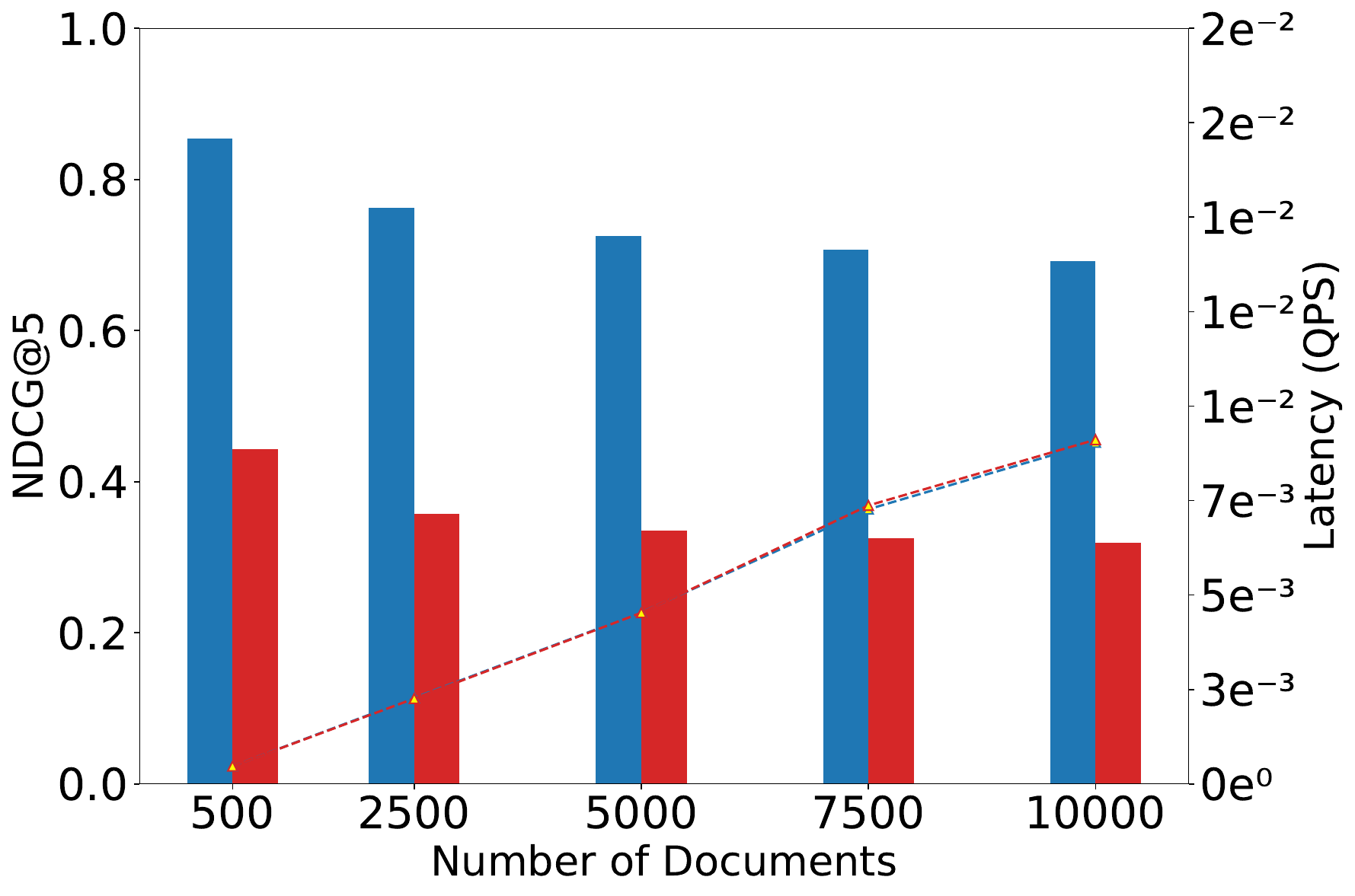}
        \caption{DSE-Qwen2 Evaluation on arXiVQA}
    \end{subfigure}
    \begin{subfigure}{0.33\textwidth}
        \includegraphics[width=\linewidth]{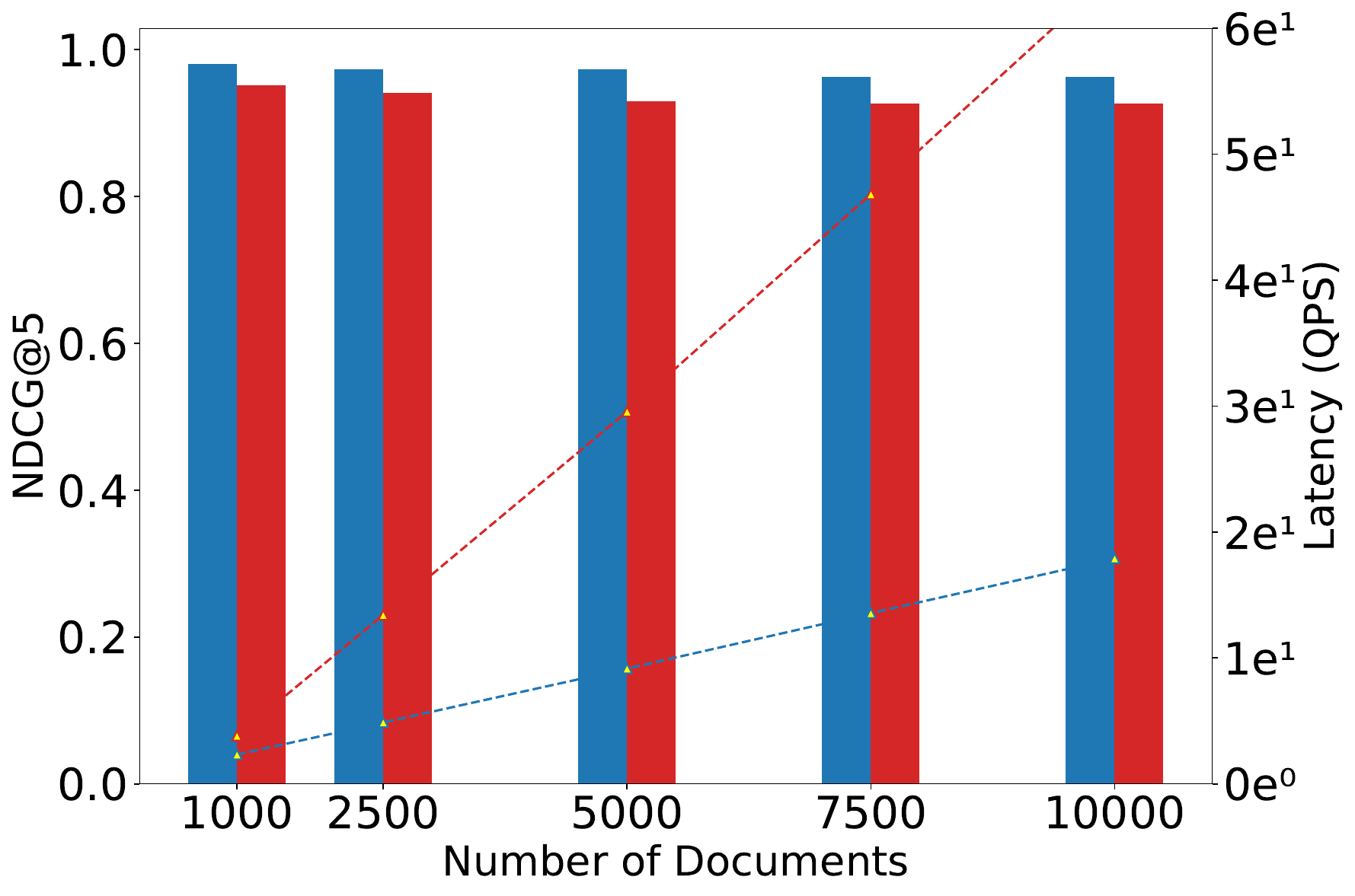}
        \caption{ColQwen2 Evaluation on Health}
    \end{subfigure}
    \hfill 
    \begin{subfigure}{0.33\textwidth}
        \includegraphics[width=\linewidth]{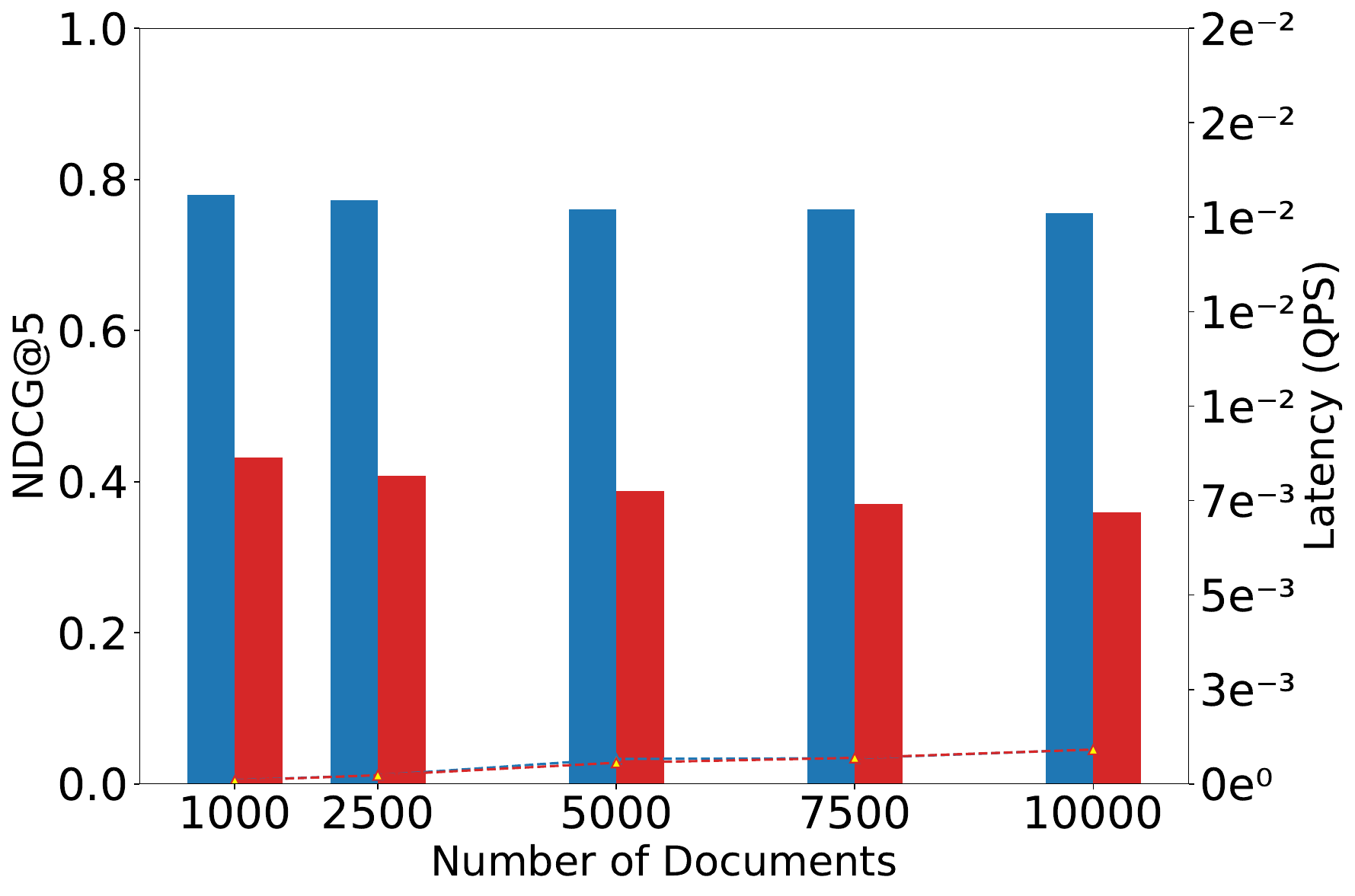}
        \caption{BiQwen2 Evaluation on Health}
    \end{subfigure}
    \hfill 
    \begin{subfigure}{0.33\textwidth}
        \includegraphics[width=\linewidth]{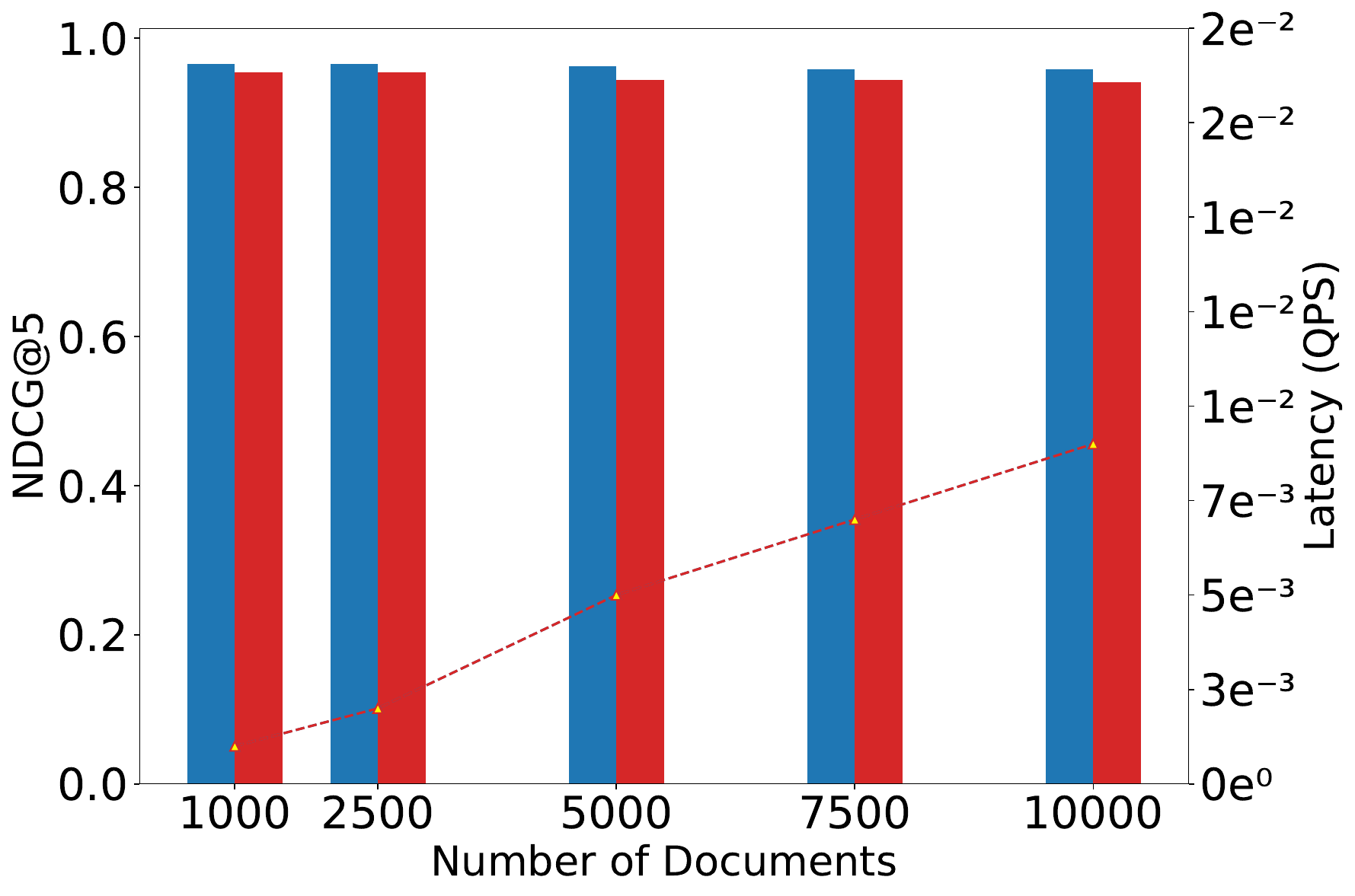}
        \caption{DSE-qwen2 Evaluation on Health}
    \end{subfigure}
    \caption{Comparison of retrieval effectiveness across various models when increasing index size for visual document and OCR-based text document retrieval; \textit{Blue} bar refers to image document indexing; \textbf{Red} bar refers to OCR-based text indexing.}
    \label{fig: 2}
    \vspace{10pt}
\end{figure*}

\vspace{1em}\noindent
\textbf{RQ1.2: Does ColPali significantly outperform the single-vector variants in terms of the effectiveness?}
One of the main contributions claimed by the original study is the proposal of late-interaction mechanism~\cite{khattab2020colbert} for document relevance estimation, as denoted in Eq.~\eqref{eq:multi}.
%
The original study shows that incorporating such multi-vector settings yields a significant performance boost compared to the single-vector baseline bi-encoder, BiPali. 
The baseline adopts the common dense text or image retrieval settings~\cite{karpukhin2020densepassageretrievalopendomain, lewis2024clip} by pooling both query token and document patch embeddings into single vectors, as described in Eq.~\eqref{eq:single} in Section~\ref{sec:preliminary}.

\paragraph{Experimental Setup} 
To verify the necessity of late interaction, we reproduce the training of the single-vector variant, BiPali, using identical training settings as ColPali.
This includes the same training epochs, datasets, and hyperparameters (See RQ 1.1).
The baseline approach was maintained to ensure consistency with ColPali while isolating the effects of the late interaction mechanism.
Additionally, we replicate another single-vector baseline, BiQwen using the Qwen2-VL as backbone like ColQwen2 in RQ 1.1.
We also re-evaluate BiPali model using the author-provided checkpoint.\footnote{\url{https://huggingface.co/vidore/bipali}}

\paragraph{Results} 
Table \ref{tab: table 1} displays the evaluation results for the single-vector variants, BiPali and BiQwen2 alongside the re-evaluated BiPali.
Similar to the gap between reported and reproduced observed in ColPali, our reproduced BiPali also shows a decrease of 2.9 nDCG@5 compared to the results re-evaluated using the provided checkpoint. 
However, this difference does not conflict with our comparison between with and without late interaction, as we controlled all other variables consistently across the experiments.
Our comparable results indicate that the single-vector baseline, BiPali performs significantly lower, with a 27.1 drop of nDCG@5, compared to one with late interaction (ColPali). Similarly, we can also observe the consistent performance drop in the replicated model BiQwen2 with a drop of 25.2 nDCG@5 compared to ColQwen2. 

\paragraph{Answer} Our reproduced baseline results justify the necessity of late interaction, which involves dynamically selecting and weighting fine-grained vectors based on the query context, substantially enhancing VDR effectiveness on the ViDoRe benchmark.

\section{Replicability}
Indexing documents with image embeddings is an innovative approach that can reduce retrieval latency by bypassing layout detection and OCR. However, the original study did not fully explore whether this approach is robust for first-stage retrieval under common scenarios. This replicability study aims to evaluate how well visual document retrieval compares to OCR-based text document embedding under two conditions typical of first-stage retrieval tasks: (1) zero-shot generalization capability and (2) larger document corpus size. 

\vspace{1em}\noindent
\textbf{RQ2.1: Does the image document embedding consistently outperform the text for first-stage retrieval?} To investigate this research question, we examine how various retrieval models perform when indexing documents either as images or as OCR-extracted text. While the original study only tested OCR text with a small-scale multi-vector retrieval model (BGE-M3~\cite{chen2024bge}, 100M-parameter), that model was not fine-tuned on more complex visual-textual elements (e.g., tables, figures). Consequently, prior findings could not conclusively show whether visual document retrieval outperforms OCR-based document retrieval.

\paragraph{Experimental Setup}  We conduct experiments to compare image-based and OCR-based indexing comprehensively, under both zero-shot and fine-tuned conditions. Two broad classes of models were evaluated: 
(1) zero-shot text-to-text retrieval model and 
(2) fine-tuned text-to-image retrieval model. 
The pure text-to-text retrieval models only handle OCR-based text document, including ColBERT-v2~\cite{santhanam2022colbertv2effectiveefficientretrieval}, BGE-M3~\cite{chen2024bge} and Jina\_ColBERT~\cite{jha2024jinacolbertv2generalpurposemultilinguallate}. 
The text-to-image retrieval models are fine-tuned on similar visual document retrieval task, including DSE-Qwen2, BiQwen2 and ColQwen2. 

\paragraph{Results and Discussion} Table~\ref{tab: table 2} summarizes the effectiveness of these models on the ViDoRe benchmark, comparing OCR-based document retrieval model with different parameter sizes. Among these, the 100 million parameter model BGE-M3 achieved an nDCG@5 of 69.4, while the retrieval effectiveness increase to 72.1 when scaling up to 7 billion. However, even this increased scale of text-to-text retrieval models does not fully compensate for the absence of visual information in retrieval effectiveness, such as arXiVQA and Gov. 

In the second block in Table~\ref{tab: table 2}, results show that fine-tuning on visual document retrieval pairs can boost the visual document retrieval performance a lot.
Surprisingly, it can also generalize to OCR-based document retrieval. 
The text-image pair retrieval model DSE-Qwen2 achieved nDCG@5 of 76.9 (for OCR text) and 84.8 (for images) on average. 
ColQwen2 with multi-vector demonstrates the highest effectiveness among all compared models with the same backbone LVLMs, achieving nDCG@5 of 77.5 (for OCR text) and 88.0 (for images). This indicates that fine-tuning on text-to-image pairs can enhance performance for both OCR-based and visual document retrieval. Finally, the results show that the visual documents have unique information, which may lost significantly from OCR conversion. We also observe the strong cross-modal generalization capability of DSE-Qwen2 and ColQwen. 
Their OCR-based document retrieval effectiveness outperforms larger text retrieval model, GTE-Qwen2.
The performance gap is particularly large for datasets with high non-textual coverage, such as ArXivQ, DocQ, and InfoQ datasets. We believe that further additional fine-tuning for text-to-text pairs might further improve results on datasets with lower textual coverage, bridging the gap on datasets rich in visual elements remains an ongoing challenge.

\paragraph{Answer} Yes, for documents that contain both textual and visual elements, visual document embedding shows more robust retrieval effectiveness compared to OCR-driven text document indexing, regardless of whether the models are fine-tuned for the visual document domain.

\vspace{20pt}
\textbf{RQ2.2: Does the image consistently outperform the OCR-based text document for first-state retrieval when the the embedding index size increases?} The aim of this section is to explore whether image-based document indexing remains viable and effective in first-stage retrieval setting that require managing larger-scale document index. Notably, the ViDoRe benchmark proposed by~\cite{faysse2024colpali} comprises at most 1,000 documents for each dataset. This scale is significantly lower compared to standard ad-hoc document retrieval settings, such as MSMARCO-passage dataset, which encompasses approximately 9 million passages. We therefore examine the effectiveness and efficiency of various fine-tuned image retrieval models, comparing image and OCR-based text embeddings at increasingly larger index sizes.
\paragraph{Experimental Setup} Because most of the datasets~\cite{li2024multimodal, mathew2021docvqa, mathew2022infographicvqa, zhu2022towards} in ViDoRe benchmark have only a limited number of documents, we focus on datasets where more than 10,000 visual documents are available in the source dataset. Specifically, we chose one academic dataset (arXivQA~\cite{li2024multimodal}) and one practical dataset (Health). 
For the arXivQA dataset, we augment the test set from the original source~\cite{li2024multimodal}, ensuring that none of the new documents appeared in the training set. Since test set of Health domain in ViDoRe benchmark is synthetic data from web, we augment the test set by using a public dataset PDFVQA~\cite{ding2023pdfvqanewdatasetrealworld} that contains medical pdf documents. 
We scale each dataset up to 100,000 documents and compare both single-vector and multi-vector models, including BiQwen2, DSE-Qwen2 and ColQwen2.

\paragraph{Results and Discussion} Figure \ref{fig: 2} shows the retrieval effectiveness and efficiency of the different models as the document corpus size increases. 
Firstly, the effectiveness of the three retrieval models on arXivQA decreases significantly as the index size increases when using both image and text document embeddings. 
Overall,  visual document retrieval demonstrates greater robustness to increased corpus size than OCR-based text document retrieval. Specifically, there is a 21\% decrease in terms of nDCG@5 for OCR-based document retrieval  when the size is expanded from 500 to 10,000 on the arXivQA dataset. On the contrary, visual document retrieval has only a 17\% nDCG@5 decrease. The health dataset exhibits only a minor decrease in effectiveness for both images and text documents. This variation is attributed to the differing complexities of the documents and queries in arXivQA versus the simplicity prevalent in the Health dataset's queries and documents.

\begin{figure}
    \centering
    \includegraphics[width=.95\linewidth]{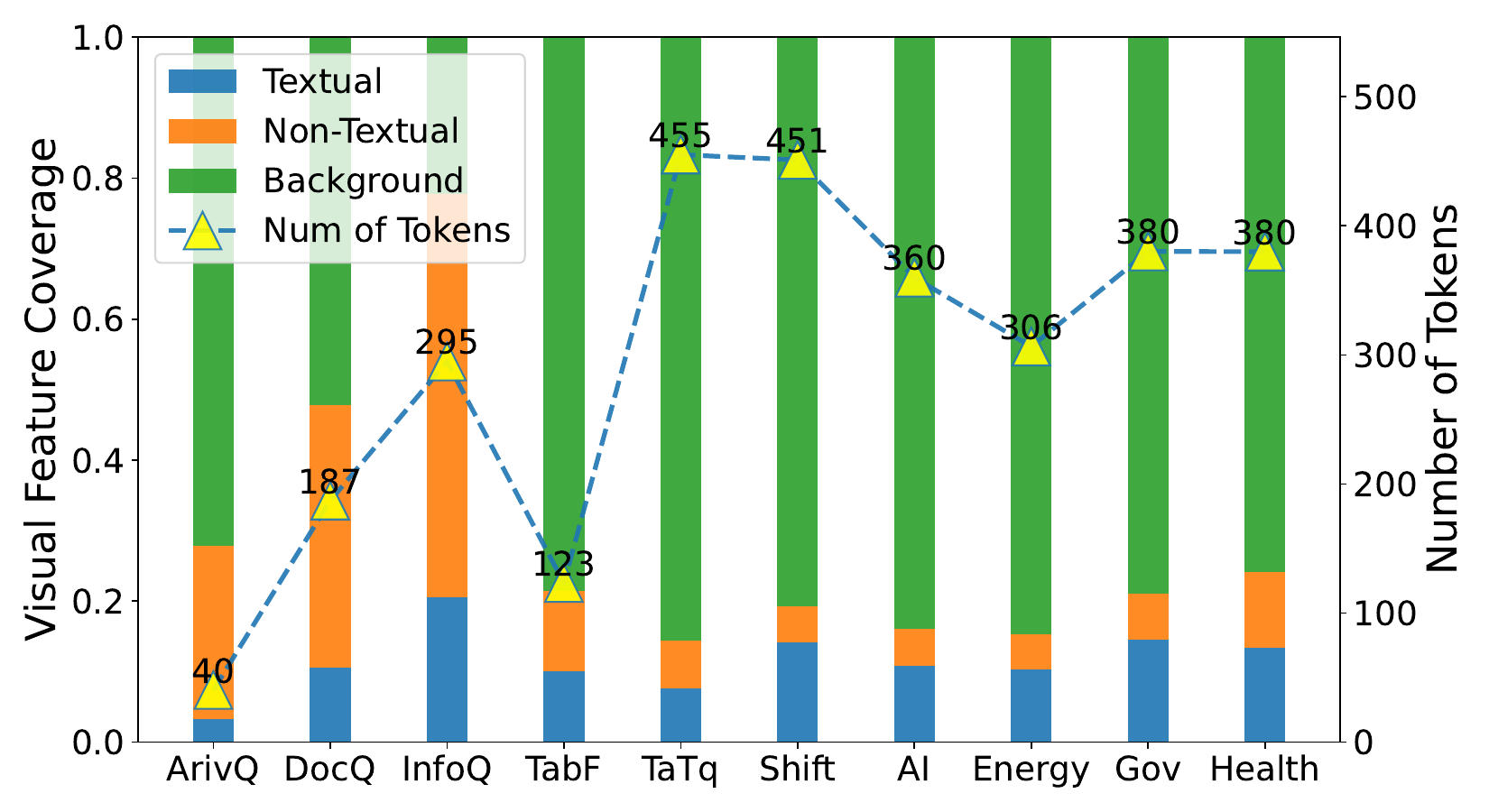}
    \caption{The visual feature distribution of datasets in ViDoRe benchmark. See Section~\ref{sec:RQ3.1} for detail.}
    \label{fig: 3}
\end{figure}

\vspace{10pt}
In terms of efficiency, the multi-vector model is less efficient than the single-vector model (BiQwen2 and DSE-Qwen2), as it requires storing more vectors (for each patch). This is due to each query and document token being represented as a vector, and the efficiency of multi-vector representations decreases as the number of tokens increases. Interestingly, the efficiency decrease for visual documents is more pronounced than for text documents in the arXivQA dataset, whereas in the Health dataset, visual documents exhibit a lower efficiency decrease than text documents. This difference is due to the higher number of text tokens in Health compared to arXivQA, despite a consistent number of patches for visual documents in both datasets. As a result, for visual documents, the number of patches—and consequently vectors—remains stable across both datasets. However, for text documents, the number of text tokens varies with the document content, leading to increased query latency as token count increases.

\paragraph{Answer} Visual document representations demonstrated a more robust effectiveness against increasing document index size compared to OCR-based text document. In terms of efficiency, single-vector models outperform multi-vector models. However, the query latency for multi-vector visual documents is lower than for multi-vector text documents when the number of patches exceeds the number of tokens.

\begin{figure*}[ht]
    \centering
    \begin{subfigure}{0.24\textwidth}
        \includegraphics[width=\linewidth]{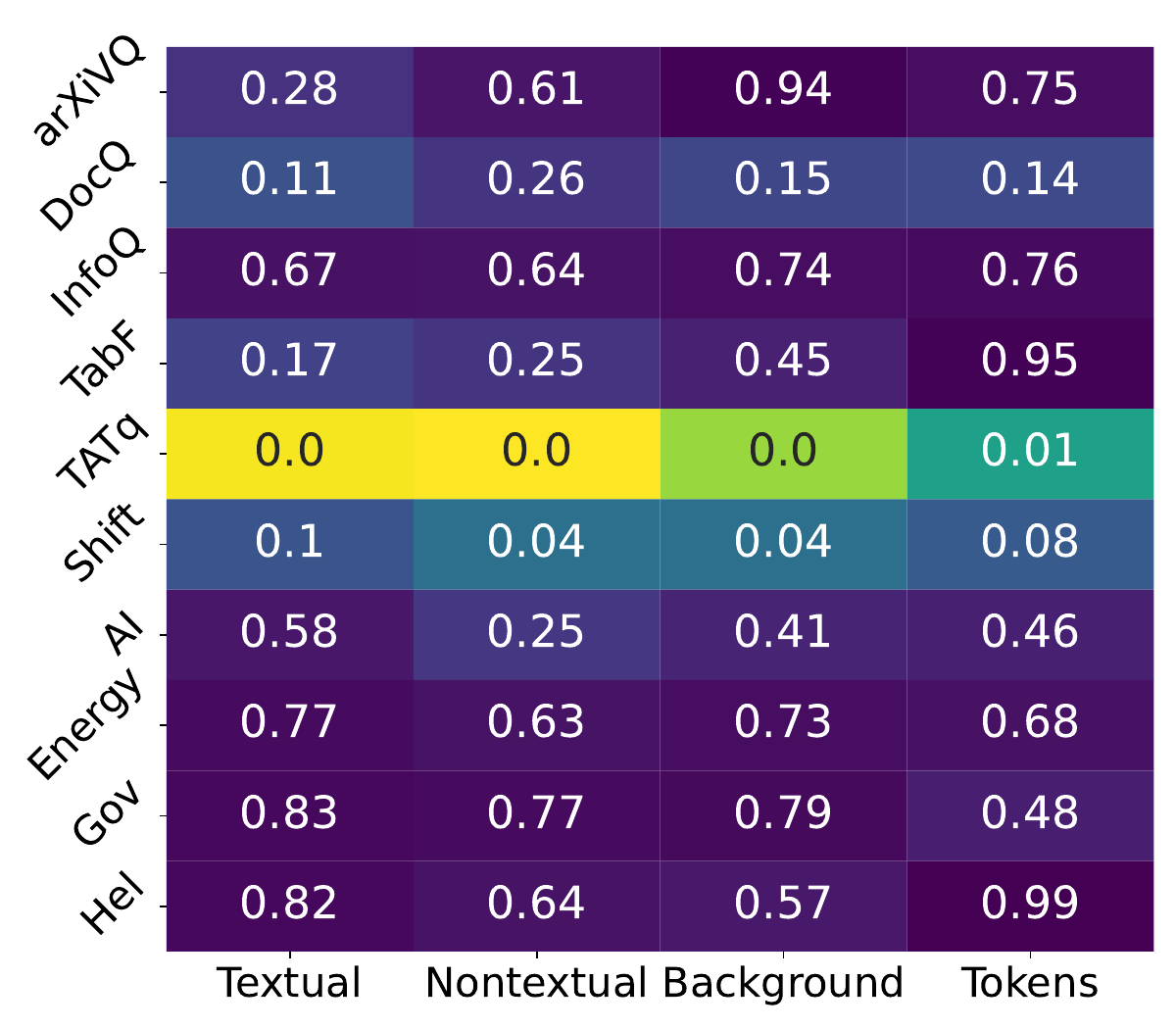}
        \caption{BiPali}
    \end{subfigure}
    \hfill
    \begin{subfigure}{0.24\textwidth}
        \includegraphics[width=\linewidth]{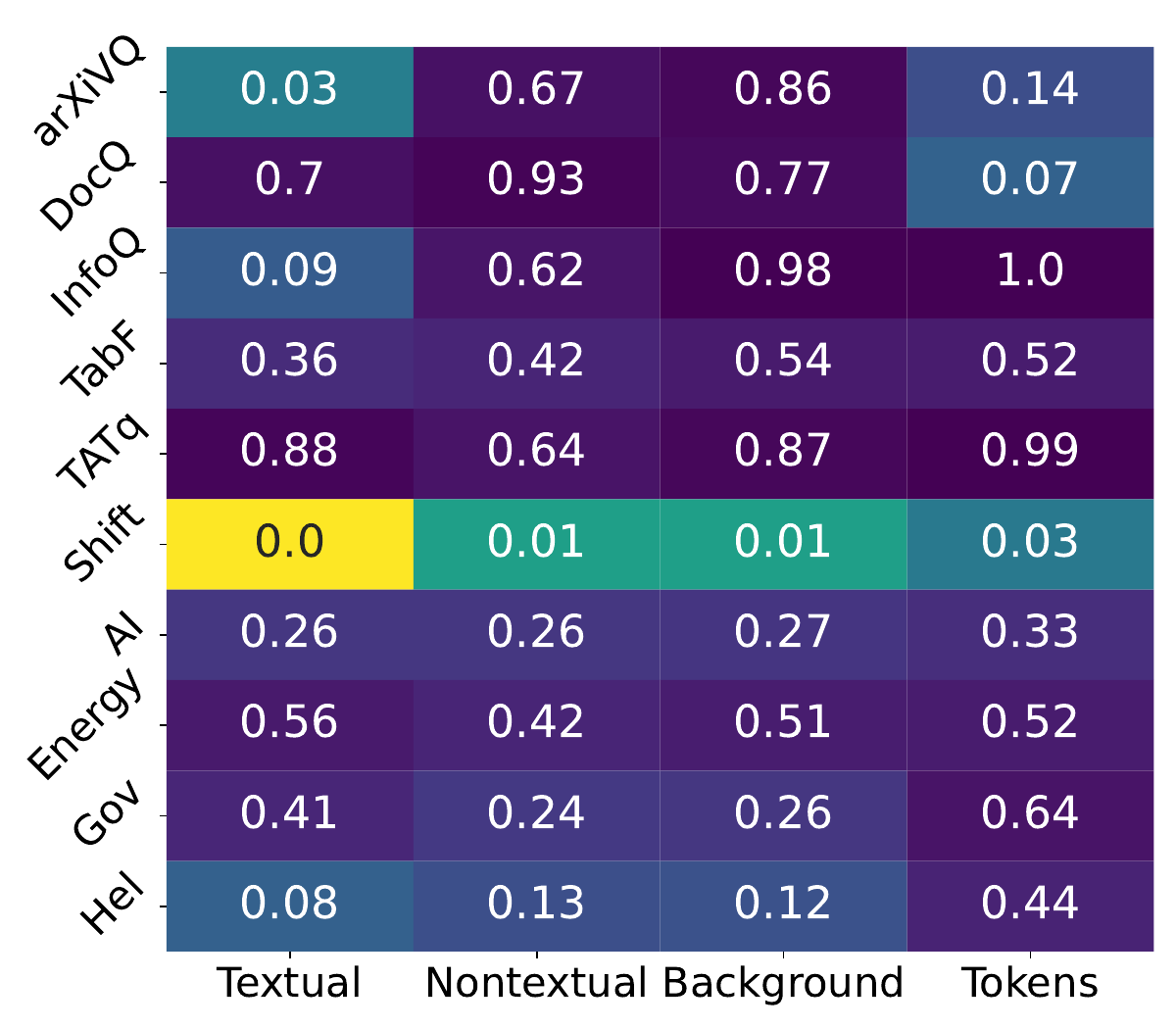}
        \caption{ColPali}
    \end{subfigure}
    \hfill
    \begin{subfigure}{0.24\textwidth}
        \includegraphics[width=\linewidth]{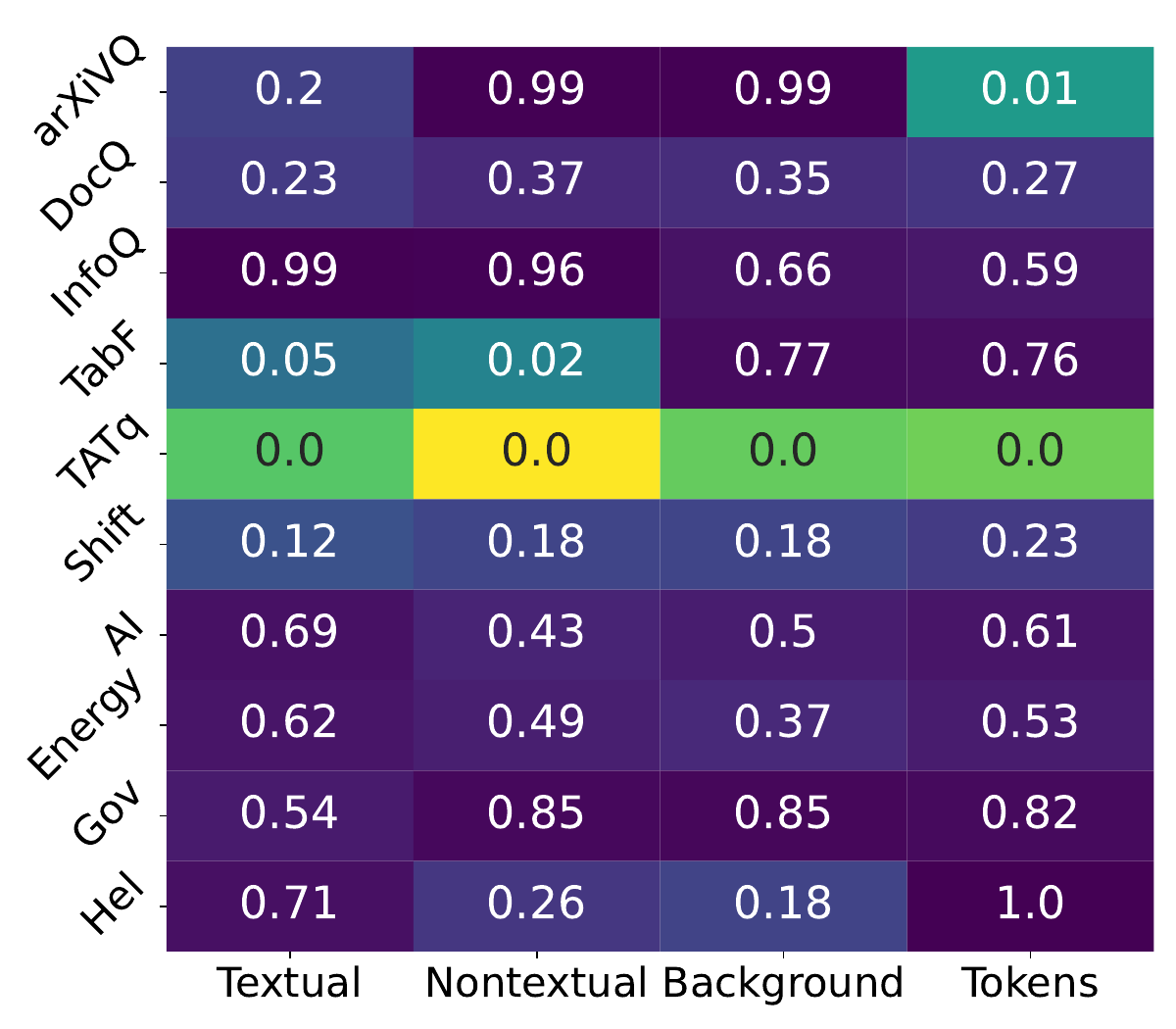}
        \caption{BiQwen2}
    \end{subfigure}
    \hfill
    \begin{subfigure}{0.24\textwidth}
        \includegraphics[width=\linewidth]{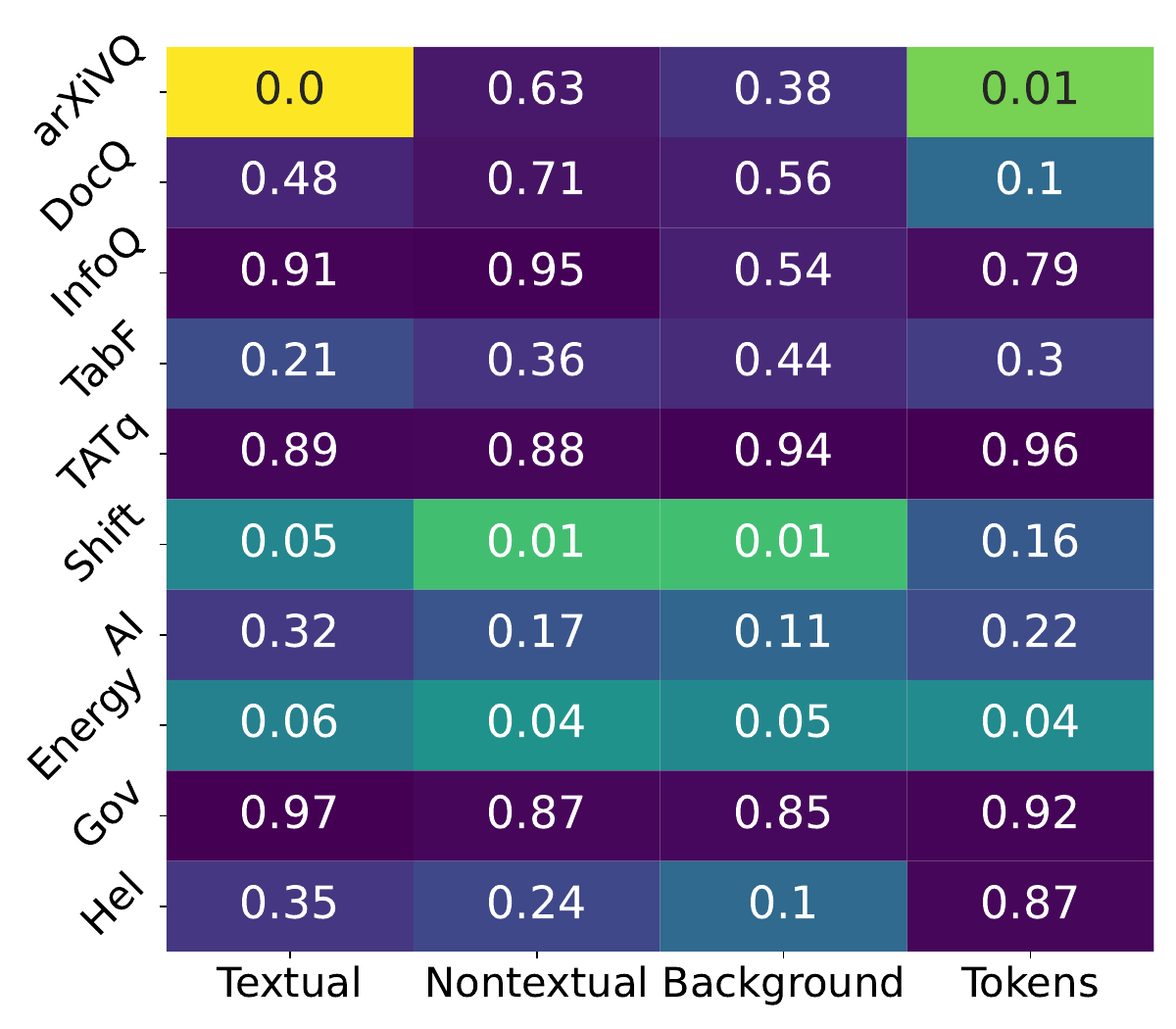}
        \caption{ColQwen2}
    \end{subfigure}
    \caption{The $p$-value of significant test of visual feature differences between retrieved and non-retrieved documents.}
    \label{fig: 4}
\end{figure*}

\section{Insights}
From our reproduction and replication studies, multi-vector models (ColPali, ColQwen2) consistently outperform single-vector models (BiQwen2, BiPali) under the same configurations (datasets, hyperparameters, and model settings). In this section, we aim to investigate the primary factors contributing to effective VDR from both document and query perspectives. We first examine from a document perspective, investigating whether there is significant visual feature influence on retrieval effectiveness (\textbf{RQ3.1}). We then shift to the query perspective, analyzing how different types of semantic matching, query and special tokens, contribute to retrieval effectiveness (\textbf{RQ3.2}), and finally investigate whether semantic query token matching primarily depends on lexical or non-lexical matching (\textbf{RQ3.3}).

\vspace{1em}\noindent
\textbf{RQ3.1:  Do significant differences exist in the visual features of retrieved image documents compared to those not retrieved?}\label{sec:RQ3.1}
To quantitatively evaluate how image features influence retrieval effectiveness, we introduce three visual feature metrics based on the coverage of pixel areas: textual coverage \(\mathcal{C}_{t}\), non-textual coverage \(\mathcal{C}_{i}\), and background coverage \(\mathcal{C}_{\varnothing}\). 

\begin{align}
    \mathcal{C}_{t} = \frac{A_{t} - A_{tbg}}{A_{total}},  \quad
    \mathcal{C}_{i} = \frac{A_{total} - A_{t} - A_{bg}}{A_{total}}, \quad
    \mathcal{C}_{\varnothing} = 1 - \mathcal{C}_{t} - \mathcal{C}_{i},
    \label{eq:visual_features}
\end{align}

where \(A_{total} \) is the total pixel area of the image. 
\(A_{t}\) is the OCR text bounding box (bbox) pixel area and \(A_{tbg}\) is the background area within the text bbox. \(A_{bg}\) is background pixel area detected based on whole image.
In addition, we measure the number of tokens in each dataset, as this factor also plays a crucial role in retrieval performance. Figure \ref{fig: 3} illustrates the distribution of these visual features across different datasets in the ViDoRe benchmark. Our goal is to determine whether relevant documents that are successfully retrieved differ significantly in these visual feature metrics compared to those that are not.

\paragraph{Experimental Setup} To answer RQ3.1, we first identify all relevant visual documents for each query and then split them into two groups: (1) Group A (True Positives): Relevant documents that the model retrieves and ranks first (recall@1=1), and (2) Group B (False Negatives): Relevant documents that the model does not rank first (recall@1=0). 
Since each group is not normally distributed, we employ Mann-Whitney U-Test to measure significance in visual features between those two groups. 
Our hypothesis is that retrieved documents (Group A) exhibit higher values in certain visual feature metrics (See Eq.~\eqref{eq:visual_features}) than Group B. We examine this hypothesis across four models: BiPali, ColPali, BiQwen2, and ColQwen2.

\paragraph{Results and Discussion} The confusion matrix in Figure \ref{fig: 4} summarizes the significance of the visual feature differences between retrieved and non-retrieved documents for both single-vector models (BiPali and BiQwen2) and multi-vector models (ColPali and ColQwen2).  We observe the most notable patterns of significance in the TATDQA, Shift, and arXiVQA datasets. On the TATDQA (finance report) dataset, Under the single-vector models BiPali and BiQwen2, true positive documents have significantly higher textual, non-textual, and background coverage than false negatives. However, these differences become less pronounced when switching to the multi-vector models ColPali and ColQwen2. On the Shift (Environmental reports) dataset, With BiPali, textual coverage and token count do not differ significantly between true positives and false negatives, though other visual features do. In contrast, using the multi-vector models amplifies differences in textual, non-textual, and background coverage, suggesting that ColPali and ColQwen2 leverage these features more effectively. On TabFQuAD (Tables) and arXiVQA (Scientific Figures), under BiQwen2, there is significantly higher textual and non-textual coverage in true positive documents for TabFQuAD, and a larger token count for arXiVQA. When extended to ColQwen2 (multi-vector), significance on TabFQuAD and TATDQA diminishes, while it increases on arXiVQA—particularly in textual coverage and token count. For other dataset (Energy and Industry) with ColPali and ColQwen2,  non-textual elements and token count become significant factors, while on Industry, background coverage shows a notable difference.
\paragraph{Answer} Yes, significant differences in visual features between retrieved and non-retrieved image documents exist in certain datasets. Notably, moving from single-vector models (BiPali, BiQwen2) to multi-vector models (ColPali, ColQwen2) tends to amplify these differences. Across eight of ten datasets, the multi-vector models exhibit stronger significance in visual features for relevant documents, suggesting that a multi-vector approach can more effectively discriminate documents based on their visual characteristics.\\

\begin{figure*}[ht]
    \centering  
    \begin{subfigure}[t]{0.38\textwidth}  
        \includegraphics[width=.95\linewidth]{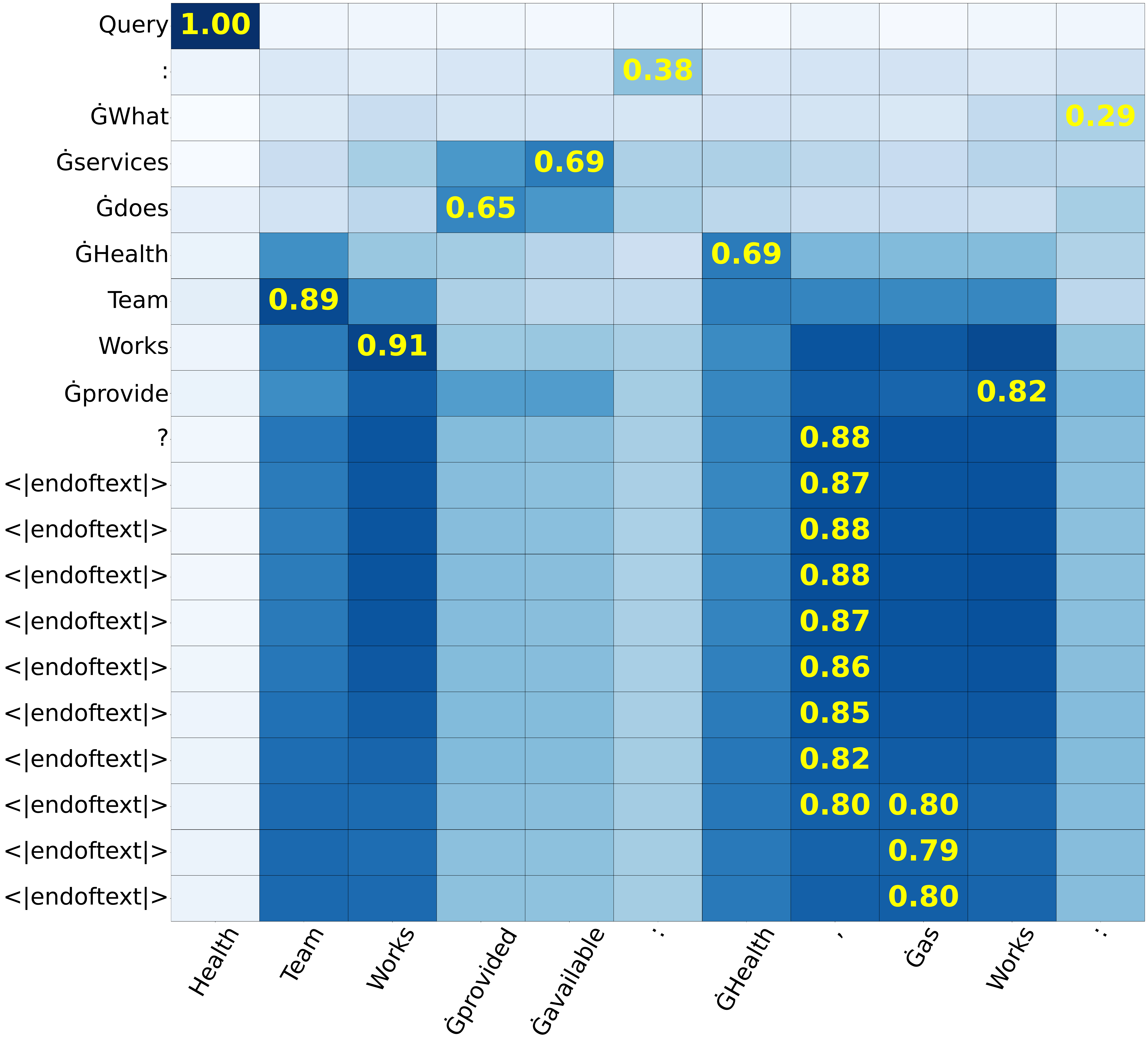}
        \caption{ColQwen2: OCR-based Text document.}
        \label{fig:text-confusion}
    \end{subfigure}
    \hfill
    \begin{subfigure}[t]{0.38\textwidth}
        \includegraphics[width=.95\linewidth]{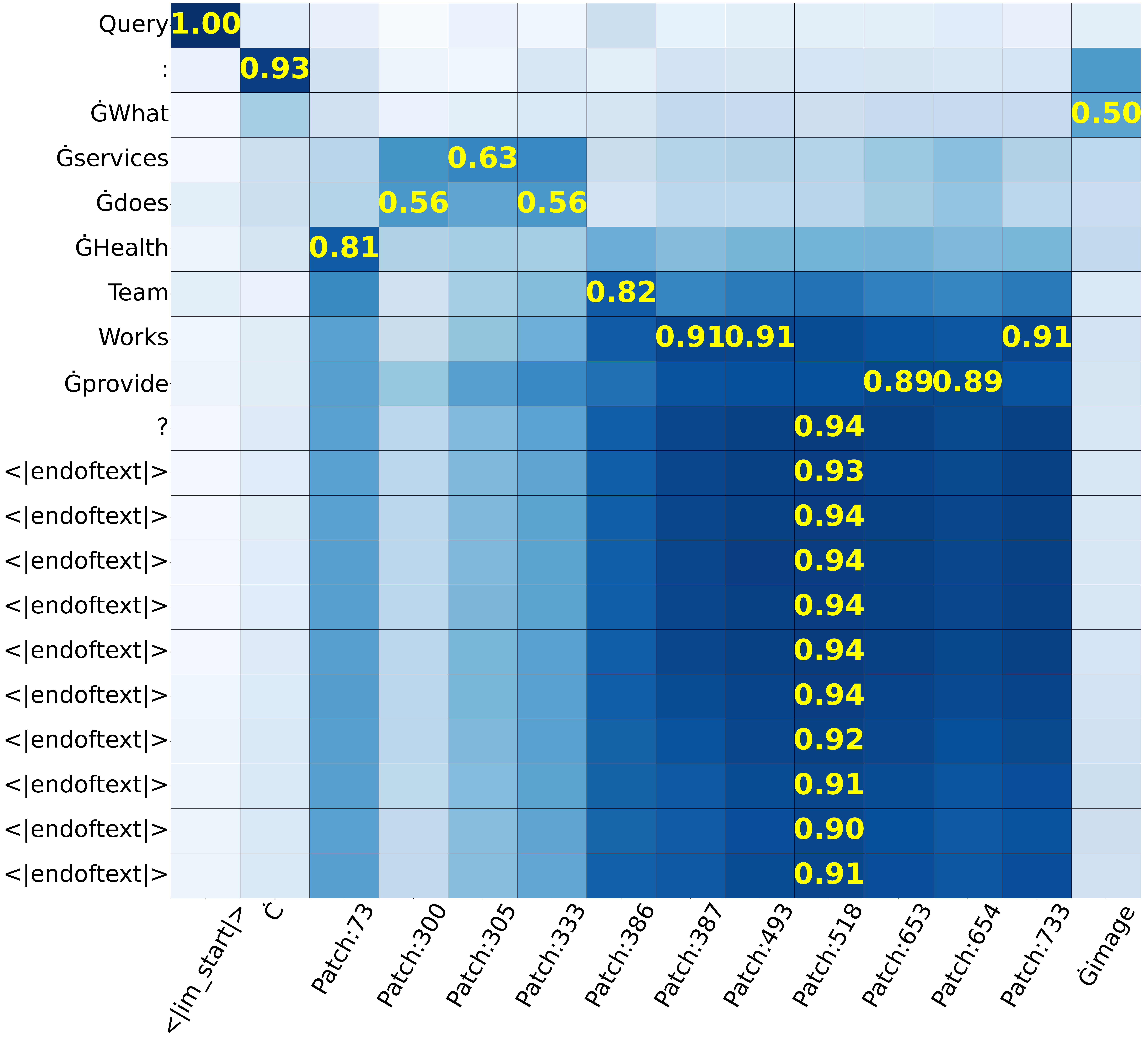}
        \caption{ColQwen2: Visual document (image).}
        \label{fig:patch-confusion}
    \end{subfigure}
    \hfill
    \begin{subfigure}{0.23\textwidth}
        \includegraphics[width=\linewidth]{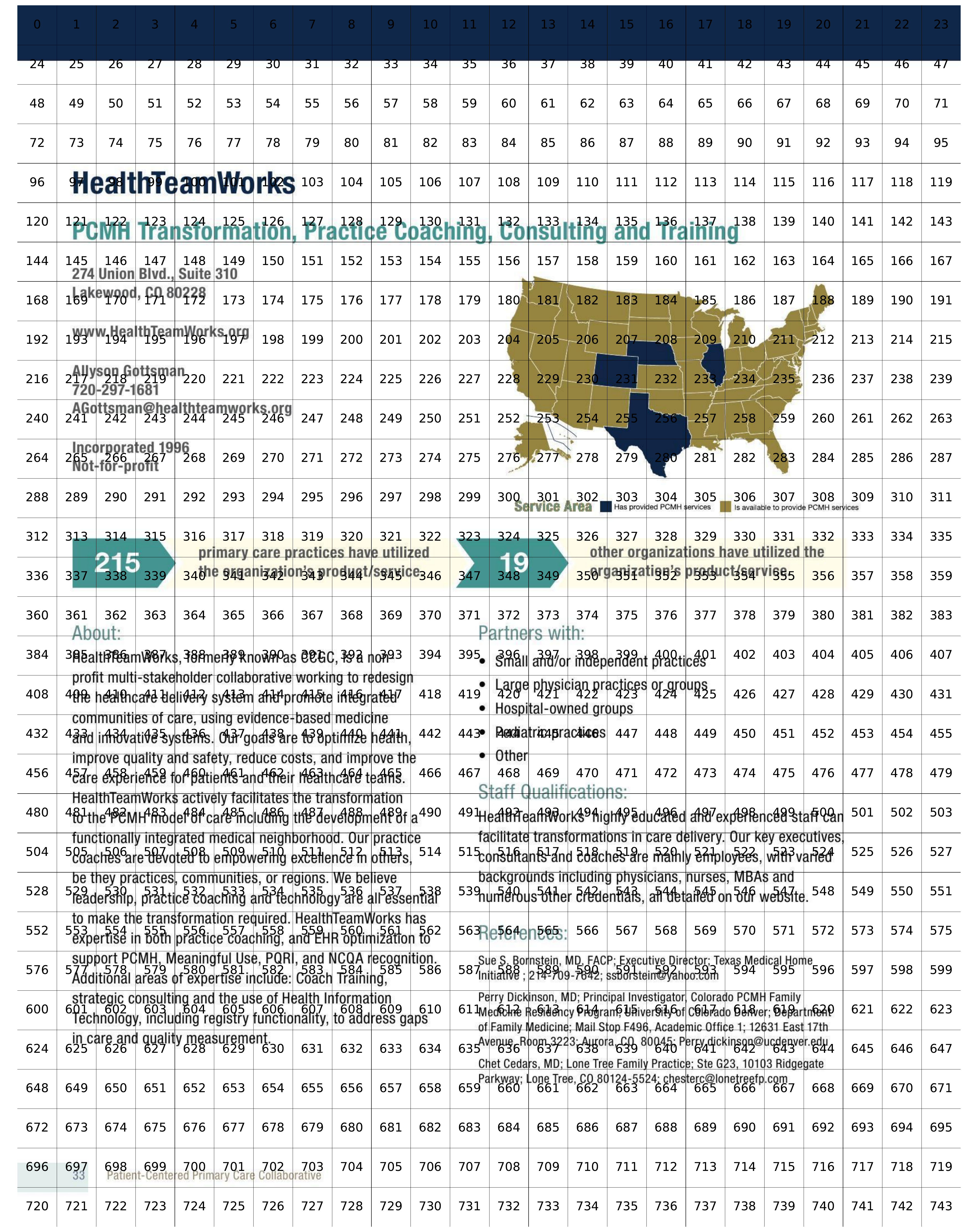}
        \vspace{10pt}
        \label{fig:similarity-map}
    \end{subfigure}
    \begin{subfigure}[t]{0.38\textwidth}  
    \includegraphics[width=.95\linewidth]{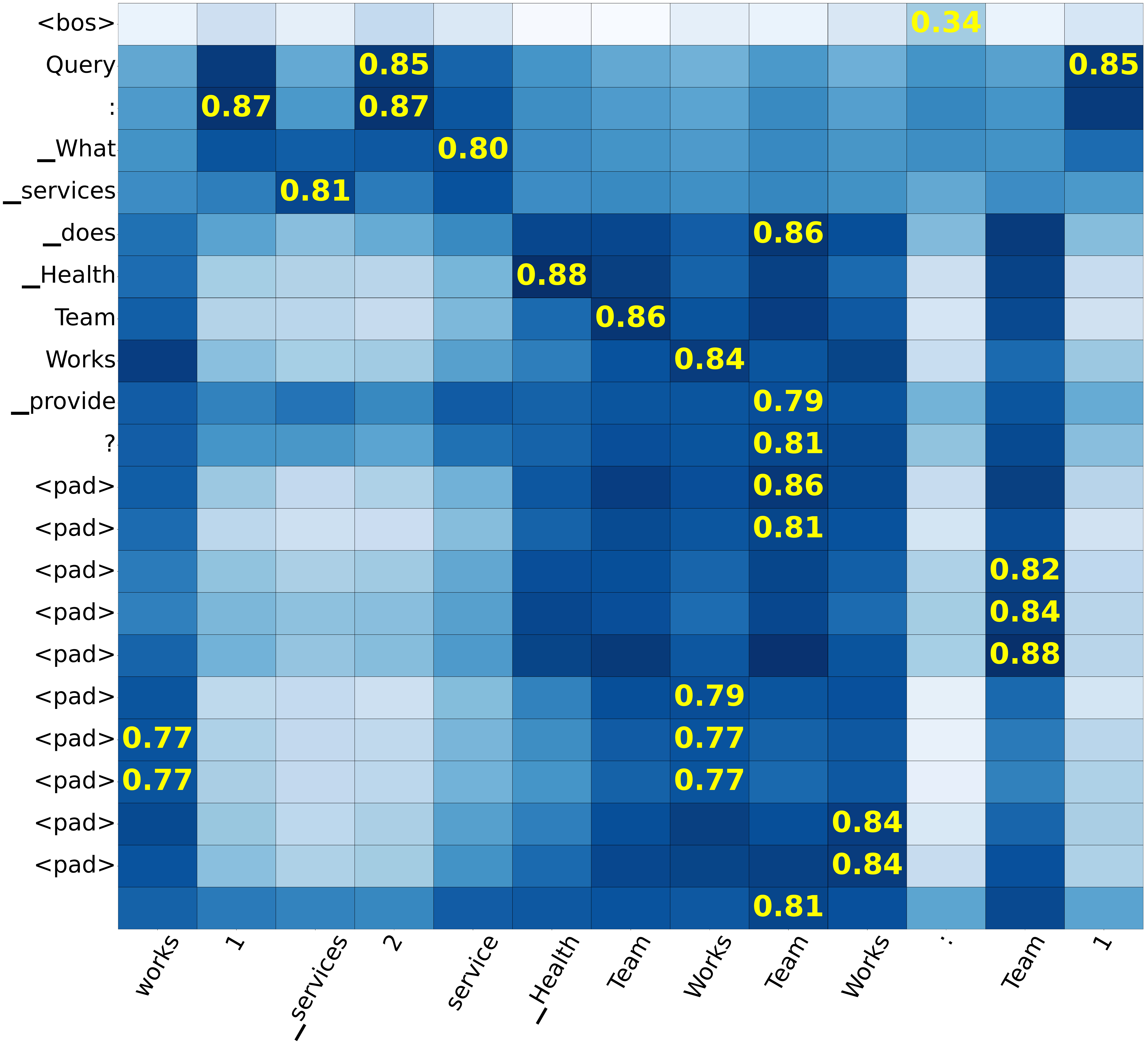}
    \caption{ColPali: OCR-based Text document}
    \end{subfigure}
    \hfill
    \begin{subfigure}[t]{0.38\textwidth}
        \includegraphics[width=.95\linewidth]{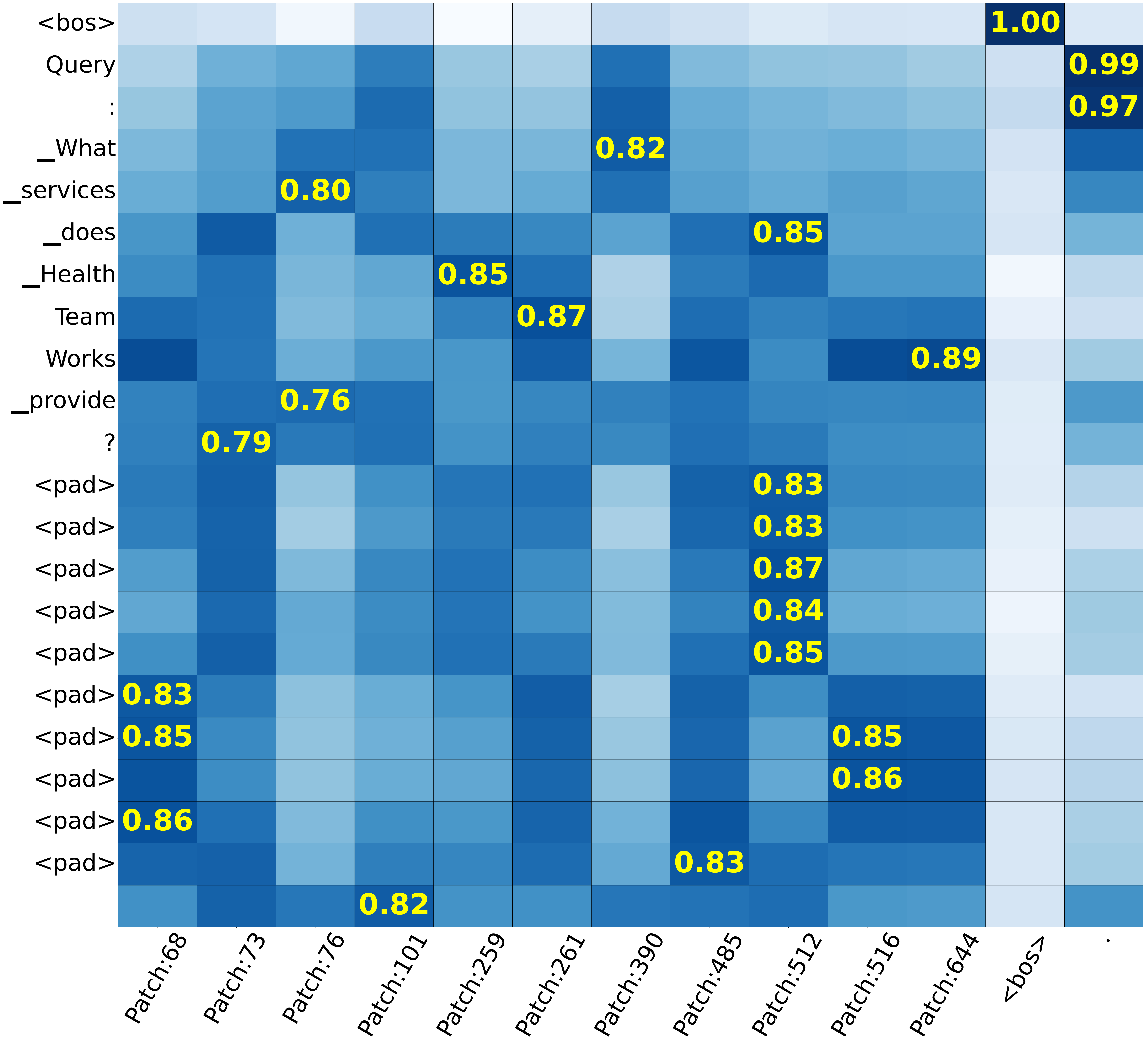}
        \caption{ColPali: Visual document (image).}
        \label{fig:patch-confusion}
    \end{subfigure}
    \hfill
    \begin{subfigure}{0.23\textwidth}
        \includegraphics[width=\linewidth]{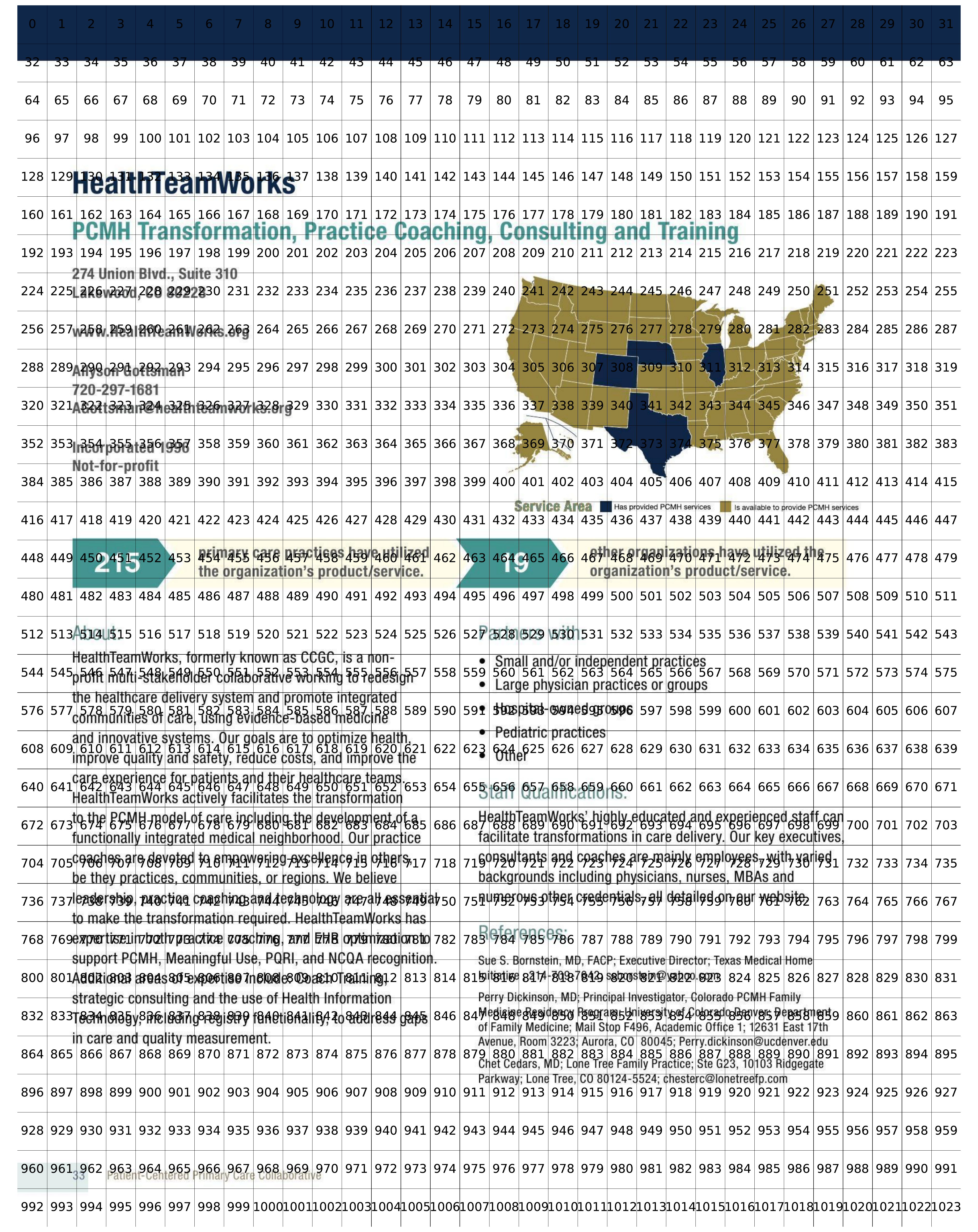}
        \vspace{10pt}
    \end{subfigure}
    \label{fig:general}

    \caption{Visualization of semantic matching types across single-vector and multi-vector models. The two images on the right display a sequence of patches generated by ColQwen (top) and ColPali (bottom), respectively. Each patch index corresponds to an index in the confusion matrix.}
    \label{fig: 5}
    \vspace{10pt}
\end{figure*}

\vspace{1em}\noindent
\textbf{RQ3.2: To What Extent Does Semantic Matching Rely on Special and Query Token Matching?}
Building on the effectiveness of multi-vector approach in discriminate documents based on their visual characteristics, we now turn to the query perspective to analysis which types of semantic matching contribute highest for retrieval results in multi-vector retrieval model. We group these into two main categories: (1) Query Token Matching (QTM) and (2) Special Token Matching (STM). QTM refers to the matching  contributions from the literal query text, whereas STM refers to the matching contribution from special tokens and prompt tokens.

Figure \ref{fig: 5} provides example illustrating multi-vector matching between query and image/text document for ColQwen2 and ColPali. From these examples, it is evident that both query text tokens and special tokens significantly influence query-document matching; both have a high similarity score. Notably, the ColPali model exhibits more variation in special token matching compared to ColQwen2, suggesting that special tokens may play a more critical role in the ColPali. This broader scoring distribution across special tokens can lead to a more nuanced matching process with document text/patch tokens. To substantiate this observation, we quantitatively assess the contributions of each matching type subsequently.

\paragraph{Experimental Setup} Evaluations were conducted using two multi-vector retrieval models, ColPali and ColQwen2, on four datasets—two in English (arXiVQA and Health) and two in French (Shift and TabF). The tests involved both text and image document, with retrieval performance measured by ndcg@5. Our experiments were specifically designed to isolate the effects of each matching type by activating only one category at a time.

\paragraph{Special vs. Query Token Matching} Table \ref{tab: table 3} shows the matching between query and image/text document. As we can seen from image document section, QTM contribute more to matching across both models for both text and image document indexing. When only special token matching is used, the effectiveness drops considerably—by an average of 20.9\% for ColPali and 19.9\% for ColQwen2 (compared to using all matching types). Conversely, query token matching is more robust, falling by only 0.1\% (ColPali) and 0.7\% (ColQwen2). These results suggest that relying exclusively on special tokens lacks sufficient contextual grounding for optimal performance. Instead, combining query tokens and special tokens appears to yield more robust retrieval. This synergy may explain why recent findings~\cite{zhuang2025documentscreenshotretrieversvulnerable} demonstrate that the multi-vector ColPali model exhibits greater robustness against certain adversarial attacks compared to the single-vector image retrieval model DSE~\cite{ma2024unifyingmultimodalretrievaldocument}. 

\paragraph{Image vs Text documents} When comparing text and image document, special token matching contributes more in text-based scenarios than with image documents. In text documents, special tokens can often align with specific phrases or keywords, providing immediate semantic overlap (e.g. align with token "Works" in above example). In contrast, in image documents, special tokens often match with image patches (e.g., white background) that have limited direct relevance, thereby reducing their impact on retrieval. 
\paragraph{Answer} Query tokens have a larger overall impact on retrieval than special tokens for both text and image queries. Special tokens alone offer insufficient context for strong performance; however, combining query tokens with special tokens can enhance robustness. Additionally, special tokens play a bigger role in text documents than in image documents due to more direct semantic overlap.

\begin{table}
    \centering
    \caption{Comparison between special tokens and query token matching in terms of nDCG@5}
    \setlength\tabcolsep{8pt}
    
    \resizebox{1\linewidth}{!}{%
    \begin{tabular}{llccc}
    \toprule
    \toprule
            Model&Datasets&  All Types &  STM&  QTM\\ 
            \midrule
   \multicolumn{5}{c}{\textit{Image document}}\\
   \midrule
  ColPali &Health & 94.0&85.5
 \textcolor{red}{\tiny{-9.0\%}}& 94.3\textcolor{red}{\tiny{0.3\%}}\\
  ColPali &arXiVQA& 82.2&59.7
 \textcolor{red}{\tiny{-27.4\%}}& 81.9\textcolor{red}{\tiny{-0.4\%}}\\
  ColPali &Shift& 76.0&50.6
 \textcolor{red}{\tiny{-33.4\%}}& 75.4\textcolor{red}{\tiny{-0.8\%}}\\
  ColPali &TabF& 84.1&72.6
 \textcolor{red}{\tiny{-13.7\%}}& 84.4\textcolor{red}{\tiny{0.3\%}}\\
  ColQwen2 &Health & 98.0&91.1
 \textcolor{red}{\tiny{-7.0\%}}& 97.2\textcolor{red}{\tiny{-0.8\%}}\\
  ColQwen2 &arXiVQA& 86.0&63.5
 \textcolor{red}{\tiny{-26.2\%}}& 85.8\textcolor{red}{\tiny{-0.2\%}}\\
  ColQwen2 &Shift& 85.7&59.2
 \textcolor{red}{\tiny{-31.0\%}}& 85.4\textcolor{red}{\tiny{-0.3\%}}\\
 ColQwen2 & TabF& 89.3& 75.4 \textcolor{red}{\tiny{-15.6\%}}& 88.2 \textcolor{red}{\tiny{-1.2}}\\
 \midrule
   \multicolumn{5}{c}{\textit{Text document}}\\
   \midrule
  ColPali &Health & 92.7& 87.3\textcolor{red}{\tiny{-5.8}}&93.5\textcolor{red}{\tiny{0.9\%}}\\ 
  ColPali &arXiVQA& 32.5& 25.8\textcolor{red}{\tiny{-20.6\%}}&31.2\textcolor{red}{\tiny{-3.9\%}}\\ 
  ColPali &Shift& 71.3& 47.9\textcolor{red}{\tiny{-32.9\%}}&71.6\textcolor{red}{\tiny{0.3\%}}\\ 
  ColPali &TabF& 77.6& 68.8\textcolor{red}{\tiny{-11.4\%}}&77.2\textcolor{red}{\tiny{-0.4\%}}\\
 ColQwen2 & Health & 95.1& 89.8\textcolor{red}{\tiny{-5.6\%}}&93.8\textcolor{red}{\tiny{-1.3\%}}\\
 ColQwen2 & arXiVQA& 38.2& 29.9\textcolor{red}{\tiny{-21.7}}&35.8\textcolor{red}{\tiny{-6.3\%}}\\
 ColQwen2 & Shift& 81.3& 58.8\textcolor{red}{\tiny{-27.6\%}}&76.7\textcolor{red}{\tiny{-5.6\%}}\\
 ColQwen2 & TabF& 82.4& 69.9\textcolor{red}{\tiny{-15.2\%}}&80.5\textcolor{red}{\tiny{-2.3\%}}\\
 \bottomrule
 \bottomrule
    \end{tabular}
    }
    \label{tab: table 3}
\end{table}

\begin{table}
    \centering
    \caption{Comparison between lexical and non-lexical matching in terms of in terms of nDCG@5.}
    \setlength\tabcolsep{3pt}
    \resizebox{1\linewidth}{!}{%
    \begin{tabular}{llccc}
    \toprule
    \toprule
 Model & Dataset& All Types & Non-lexical QTM& 
 Lexical QTM\\
 \midrule
 \multicolumn{5}{c}{\textit{Text document}}\\ 
 \midrule
  ColPali &Health & 92.7& 43.1\textcolor{red}{\tiny{-53.5\%}}& 60.2\textcolor{red}{\tiny{-35.1\%}}\\ 
  ColQwen2 &Health & 95.1& 43.4\textcolor{red}{\tiny{-54.4\%}}& 59.4\textcolor{red}{\tiny{-37.5\%}}\\ 
  ColPali &arXiVQA& 32.5& 27.1\textcolor{red}{\tiny{-16.6\%}}& 10.4\textcolor{red}{\tiny{-68.0\%}}\\ 
  ColQwen2 &arXiVQA& 38.2& 25.7\textcolor{red}{\tiny{-32.7\%}}& 12.8\textcolor{red}{\tiny{-66.5\%}}\\ 
  ColPali &Shift& 71.3& 16.7\textcolor{red}{\tiny{-76.6\%}}& 43.6\textcolor{red}{\tiny{-38.9\%}}\\ 
  ColQwen2 &Shift& 81.3& 12.1\textcolor{red}{\tiny{-85.1\%}}& 48.8\textcolor{red}{\tiny{-39.9\%}}\\ 
  ColPali &TabF& 77.6& 65.9\textcolor{red}{\tiny{-15.0\%}}& 38.6\textcolor{red}{\tiny{-50.3\%}}\\ 
  ColQwen2 &TabF& 82.4& 61.2\textcolor{red}{\tiny{-25.7\%}}& 45.1\textcolor{red}{\tiny{-45.3\%}}\\
  \bottomrule
  \bottomrule
    \end{tabular}
    }
    \label{tab: table 4}
\end{table}

\vspace{1em}\noindent
\header{RQ3.3: Does Query Token Matching Depend on Lexical or Non-Lexical Matching?}
We now further dissect query token matching, investigating whether it is primarily driven by lexical or non-lexical factors. Previous research~\cite{wang2023repro} highlights a strong preference for lexical similarity. Our objective is to determine to what extent lexical signals govern query-token matching in both image-based and OCR-driven text-based retrieval.

\paragraph{Experimental Setup} Using the same setup as in RQ3.2, we focus on distinguishing lexical from non-lexical query token matches. Because direct lexical matching between text queries and image patches is inherently limited, our analysis for image patches is largely qualitative (case studies). However, we do perform a quantitative breakdown for text-document retrieval, separating out lexical from non-lexical matches.

\paragraph{Case study} As depicted in Figure \ref{fig: 5}, shows that image-document matching often hinges on identical or near-identical tokens within a document patch or its adjacent patches.  For instance, in the query  \textit{what services does health team works provide} for example,  we observe: the term \textit{"what"} aligning with a specific character patch; \textit{"services"} and \textit{"does"} mapping to 305 patch containing \textit{service};  query token \textit{"health"} corresponds to the 73 patch (above the patch with “health”); \textit{"team"} is associated with the 75 patch (above the patch with \textit{"team"}); \textit{"works"} correlates with the 387 patch with token works; \textit{"provide"} matches both the 446 patch (\textit{"ctices"}) and the 653 patch (white background).  These patterns, also observed using ColPali, suggest that VDR benefits from the matching of identical tokens and surround patches. These patterns, observed suggest that lexical overlaps (even partial ones) in or near the relevant patch enhance image retrieval performance.

\paragraph{Lexical Matching vs. Non-Lexical Matching} We next conducted a quantitative evaluation of lexical vs. non-lexical matching for OCR-driven text across various datasets, as shown in Table \ref{tab: table 4}. In the Datasets with fewer tokens, such as arXiVQA and TabF, rely more heavily on non-lexical matching. In contrast, datasets with a larger number of tokens, such as Health and Shift, depend more heavily on lexical matching. Thus, reducing lexical matching impacts retrieval effectiveness more than reducing non-lexical matching. For instance, ColQwen2's performance drops by 85.1\% on larger token dataset Shift when using only non-lexical QTM, and 39.9\% on lexical QTM. This pattern suggests that multi-vector models may switch to non-lexical matching when lexical tokens are unavailable.

\paragraph{Answer} For image documents, query token matching typically involves lexical matches when possible, supplemented by nearby patches. For OCR-driven text documents, while lexical matching dominates in datasets with abundant text, there is a shift towards non-lexical matching in scenarios where lexical tokens are insufficient or absent. 

\section{Conclusions}
In this study, we first reproduce and validate the core techniques of ColPali in the context of visual document retrieval, especially the importance of late interaction. Second, we examine ColPali under designed practical conditions, observing that fine-tuning on visual documents brings extra generalization to OCR-based document retrieval. However, the performance degrades as the corpus size increases. Finally, we probe ColPali from both document and query perspectives, showing that the model is sensitive to document variance. This is also evident by diverse query-matched patches within the document.

\section*{Acknowledgments} We thank all reviewers for their feedback. This research was supported by the China Scholarship Council under grant number 202208410053 and project VI.Vidi.223.166 of the NWO Talent Programme (partly) financed by the Dutch Research Council (NWO).
The views expressed in this paper are those of the authors and do not necessarily reflect the views of their institutions or sponsors. 

\bibliographystyle{ACM-Reference-Format}
\bibliography{reference}

\end{document}